\documentclass[useAMS,usenatbib]{mn2e}
\usepackage[dvips]{graphicx}
\usepackage{txfonts}
\usepackage{amssymb}
\usepackage{indentfirst}
\usepackage[T1]{fontenc}
\usepackage[latin1]{inputenc} 
\begin{document}

\title[Unresolved X-ray emission in nearby galaxies]{Unresolved and diffuse components of X-ray emission and X/K luminosity ratios in nearby early-type and late-type galaxies}

\author[\'A. Bogd\'an \& M. Gilfanov]{\'A. Bogd\'an$^{1,2}$\thanks{E-mail:
abogdan@cfa.harvard.edu (\'AB); gilfanov@mpa-garching.mpg.de (MG)} 
and
M. Gilfanov$^{2,3}$\footnotemark[1]\\
$^{1}$Smithsonian Astrophysical Observatory, 60 Garden Street,
Cambridge, MA 02138, USA \\
$^{2}$Max-Planck-Institut f\"ur Astrophysik, Karl-Schwarzschild-Str.1,
85741 Garching bei M\"unchen, Germany\\
$^{3}$Space Research Institute, Russian Academy of Sciences, Profsoyuznaya
84/32, 117997 Moscow, Russia}

\date{}

\maketitle

\begin{abstract}
We explore the nature of  unresolved X-ray emission in a broad sample of  galaxies of all morphological types  based on archival \textit{Chandra} data. After removing bright compact sources, we study  $L_X/L_K$ luminosity ratios of unresolved emission, and compare them with the Solar neighborhood values. We conclude that unresolved emission is determined by four main components, three of which were known before: (i) The population of faint unresolved  sources associated with old stellar population. In early-type galaxies, their $2-10$ keV band luminosity scales with the stellar mass with $L_X/L_K=(3.1\pm 0.9) \times 10^{27} \ \rm{erg \ s^{-1} \ L^{-1}_{K,\odot}}$;
(ii) The ISM with  $kT\sim 0.2-0.8$ keV present in galaxies of all types. Because of the large dispersion in the gas content of galaxies, the size of our sample is insufficient to obtain reliable scaling law for this component; (iii) The population of unresolved young stars and young stellar objects in late-type galaxies. Their $2-10$ keV band luminosity  scales with the star-formation rate with  $L_X/$SFR $ \approx (1.7\pm0.9) \times10^{38} \ \mathrm{(erg/s)/(M_{\odot}/yr)}$; (iv) In four old and massive Virgo ellipticals  (M49, M60, M84, NGC4636) we find anomalously high X-ray emission in the $2-10$ keV band.  Its presence has not been recognized before and its nature is unclear.  Although it appears to be stronger in galaxies having stronger ISM component, its existence cannot be explained in terms of  an extrapolation of the warm ISM spectrum. Association with Virgo cluster of galaxies suggests that the excess emission may be due to  intracluster gas accreted in the gravitational well of a massive galaxy. We investigate this and other possibilities. 
\end{abstract}


\section{Introduction}
The X-ray appearance of most nearby ($D\lesssim30$ Mpc) galaxies is determined by bright X-ray binaries \citep{fabbiano06}, such as high-mass X-ray binaries (HMXBs) and low-mass X-ray binaries (LMXBs). Because of their high X-ray luminosity ($L_X\sim10^{35}-10^{39} \ \rm{erg \ s^{-1}}$) notable fraction of them can be resolved with moderately deep \textit{Chandra} exposures, and hence hence they can be studied in full particulars \citep[e.g.][]{grimm03,gilfanov04, mineo11}. Underneath bright X-ray binaries diffuse emission is present in all types of galaxies. Similarly to the Galactic Ridge X-ray emission \citep{revnivtsev06}, a part of this emission originates from the collective emission of faint  ($L_X\sim10^{27}-10^{35} \ \rm{erg \ s^{-1}}$) compact X-ray sources, mostly from active binaries (ABs), cataclysmic variables (CVs), young stars, and young stellar objects (YSOs). Based on the analysis of a few nearby early-type galaxies, it was suggested that the overall luminosity of ABs and CVs is tightly correlated with the stellar mass of the galaxy  \citep[e.g.][]{pellegrini94,bogdan, revnivtsev08}. Additionally, in all types of galaxies truly diffuse emission from warm ($kT \lesssim 1 $ keV) ionized interstellar  medium (ISM) is also present, whose measure, among others, depends  on the mass of the host galaxy, albeit with large dispersion \citep{osullivan01,mathews03}. The total observed unresolved emission is the combination of these, and possibly other, components.

\begin{table*}
\caption{The list of early-type and late-type galaxies studied in this paper.}
\begin{minipage}{18cm}
\renewcommand{\arraystretch}{1.3}
\centering
\begin{tabular}{c c c c c c c c c c c}
\hline 
Name & Distance &  $L_{K} $               &  $N_{H}$ & Morphological &  SFR     & Age & $ T_{\mathrm{obs}} $ & $ T_{\mathrm{filt}} $ & $L_{\mathrm{lim}}$            & $R$ \\ 
     & (Mpc)    &($\mathrm{L_{K,\odot}}$) &(cm$^{-2}$) & type        & ($\mathrm{M_{\odot} \ yr^{-1}}$)  & (Gyrs) & (ks)  & (ks)                  &  ($ \mathrm{erg \ s^{-1}} $)  & ($\arcsec$)   \\ 
     &   (1)    &            (2)           &   (3)      &     (4)     &      (5)                 &   (6) &  (7)                  &   (8)                         & (9) & (10) \\
\hline 
M51         & $ 8.0^a   $ & $ 6.4 \times 10^{10} $& $ 1.6 \times 10^{20} $ & SAbc         & 3.9 & -- & $  90.9 $ & $  80.8 $ & $ 9 \times 10^{36} $ &  175 \\   
M74         & $ 7.3^a   $ & $ 1.6 \times 10^{10} $& $ 4.8 \times 10^{20} $ & SA(s)c       & 1.1 & -- & $ 104.4 $ & $  84.7 $ & $ 7 \times 10^{36} $ &  140 \\   
M81         & $ 3.6^b   $ & $ 4.8 \times 10^{10} $& $ 4.1 \times 10^{20} $ & SA(s)ab      & 0.4& -- & $ 239.1 $ & $ 202.1 $ & $ 8 \times 10^{35} $ &  250 \\
M83         & $ 4.5^a   $ & $ 3.9 \times 10^{10} $& $ 3.9 \times 10^{20} $ & SAB(s)c      & 2.8 & -- & $  61.6 $ & $  52.1 $ & $ 4 \times 10^{36} $ &  200 \\
M94         & $ 4.7^a   $ & $ 3.3 \times 10^{10} $& $ 1.4 \times 10^{20} $ & (R)SA(r)ab   & 1.2& -- & $  76.9 $ & $  69.5 $ & $ 4 \times 10^{36} $ &  164 \\
M95         & $10.1^a   $ & $ 3.3 \times 10^{10} $& $ 2.9 \times 10^{20} $ & SB(r)b       & 1.1 & -- & $ 120.1 $ & $  99.6 $ & $ 1 \times 10^{37} $ &   88 \\
M101        & $ 7.4^a   $ & $ 3.4 \times 10^{10} $& $ 1.2 \times 10^{20} $ & SAB(rs)cd    &1.3& -- & $1070.6 $ & $ 833.3 $ & $ 6 \times 10^{35} $ &  167 \\
NGC2403     & $ 3.2^b   $ & $ 5.0 \times 10^{9}  $& $ 4.2 \times 10^{20} $ & SAB(s)cd    &0.4& --& $ 224.0 $ & $ 184.2 $ & $ 7 \times 10^{35} $ &  140 \\
NGC3077     & $ 3.8^a   $ & $ 2.6 \times 10^{9}  $& $ 4.0 \times 10^{20} $ & I0 pec       & 0.3 &-- & $  54.1 $ & $  42.1 $ & $ 3 \times 10^{36} $ &  77.5 \\ 
NGC4214     & $ 2.9^a   $ & $ 6.5 \times 10^{8}  $& $ 1.5 \times 10^{20} $ & IAB(s)m      & 0.2 & -- &$  83.4 $ & $  53.6 $ & $ 1 \times 10^{35} $ &   70\\
NGC4449     & $ 4.2^a   $ & $ 3.4 \times 10^{9}  $& $ 1.4 \times 10^{20} $ & IBm          & 0.4 & -- & $ 102.1 $ & $  97.0 $ & $ 2 \times 10^{35} $ &   95\\
M31 bulge   & $ 0.78^a  $ & $ 3.7 \times 10^{10} $& $  6.7 \times 10^{20}$ &  SA(s)b      &  --  &  5.1$^f$ & $ 180.1 $ & $ 143.7 $& $ 2 \times 10^{35} $ & $ 720 $  \\   
M32         & $ 0.805^b $ & $ 8.5 \times 10^{8}  $& $  6.3 \times 10^{20}$ &  cE2         &  --  &  3.8$^f$ & $ 178.8 $ & $ 172.2 $& $ 1 \times 10^{34} $& $  90 $  \\   
M49         & $ 16.3^c  $ & $ 3.0 \times 10^{11} $& $  1.7 \times 10^{20}$ &  E2          &  --  &  8.5$^f$ &$ 60.9 $  & $ 55.7  $& $ 5\times 10^{37}$& $  140 $  \\ 
M60         & $ 16.8^c  $ & $ 2.2 \times 10^{11}$ & $  2.2 \times 10^{20}$ &  E2          &  --  &  11.0$^f$ & $ 109.3 $ & $  90.5 $& $ 2 \times 10^{37} $ & $ 125 $  \\
M84         & $ 18.4^c  $ & $ 1.6 \times 10^{11}$ & $  2.6 \times 10^{20}$ &  E1          &  --  & 11.8$^f$ &$ 117.0 $ & $ 113.8 $& $ 3 \times 10^{37} $ & $ 100 $  \\
M89         & $ 15.3^c  $ & $ 7.0 \times 10^{10}$ & $  2.6 \times 10^{20}$ &  E0-         &  --  & 9.6$^f$ & $ 55.1 $  & $  51.8 $& $ 4 \times10^{37} $ & $ 80  $ \\
M105        & $ 9.8^d   $ & $  4.1 \times 10^{10}$& $  2.8 \times 10^{20}$ &  E1          &  --  & 9.3$^f$  &$ 341.4 $ & $ 314.0 $& $ 2 \times 10^{36} $ & $  90 $  \\
NGC1291     & $ 8.9^e   $ & $  6.3 \times 10^{10}$& $  2.1 \times 10^{20}$ &  (R)SB(s)0/a &  --  & -- &$ 76.7 $  & $ 51.2  $& $ 1 \times 10^{37} $ & $ 130 $  \\
NGC3377     & $ 11.2^c  $ & $  2.0 \times 10^{10}$& $  2.9 \times 10^{20}$ &  E5-6        &  --  & 4.1$^f$ & $ 40.2 $  & $ 34.0 $ & $ 3 \times 10^{37} $ & $  79 $  \\
NGC3585     & $ 20.0^c  $ & $  1.5 \times 10^{11}$& $  5.6 \times 10^{20}$ &  E7/S0       &  --  & 3.1$^f$ &$ 95.9 $  & $ 89.1 $ & $ 3 \times 10^{37} $ & $ 180 $  \\
NGC4278     & $ 16.1^c  $ & $  5.5 \times 10^{10}$& $  1.8 \times 10^{20}$ &  E1-2        &  --  & 10.7$^f$& $ 467.7 $ & $ 443.0 $& $ 4 \times 10^{36} $ & $ 110 $  \\ 
NGC4365     & $ 20.4^c  $ & $  1.1 \times 10^{11}$& $  1.6 \times 10^{20}$ &  E3          &  --  &  3.6$^g$ &$ 198.3 $ & $ 181.4 $& $ 2 \times 10^{37} $ & $  80 $  \\
NGC4526     & $ 16.9^c  $ & $  9.6 \times 10^{10}$& $  1.7 \times 10^{20}$ &$\rm{SAB(s)0^0}$ &-- & 1.7$^h$ & $ 44.1 $  & $ 34.5  $& $ 5 \times 10^{37} $ & $  84 $  \\
NGC4636     & $ 14.7^c  $ & $  8.1 \times 10^{10}$& $  1.8 \times 10^{20}$ &  E/S0\_1     &  --  & 10.3$^i$ & $ 212.5 $ & $ 202.1 $& $ 4 \times 10^{37} $ & $ 102 $  \\
NGC4697     & $ 11.8^c  $ & $  5.1 \times 10^{10}$& $  2.1 \times 10^{20}$ &  E6          &  --  & 8.2$^f$ & $ 195.6 $ & $ 162.0 $& $ 8 \times 10^{36} $ & $  95 $  \\
\hline \\
\end{tabular} 
\end{minipage}
\textit{Note.} Columns are as follows. (1) References are:  $^a$ \citet{karachentsev04}  -- $^b$ \citet{freedman01} -- $^c$ \citet{leonard02} -- $^a$ \citet{stanek,macri} -- $^b$ \citet{m32distance} -- $^c$ \citet{ngc4278distance} -- $^d $ \citet{m105distance} -- $^e$ \citet{ngc1291distance}. (2) Total near-infrared luminosity of the elliptic region described in column (8). (3) Galactic absorption \citep{dickey90}. (4) Taken from NED (http://nedwww.ipac.caltech.edu/). (5) Star-formation rate for late-type galaxies within the same ellipse. (6) Age of the stellar population for early-type galaxies. For NGC1291 no age reference has been found. References are: $^f$ \citet{terlevich02} -- $^g$ \citet{denicolo05} -- $^h$ \citet{gallagher08} -- $^i$ \citet{sanchez06}.  (7) and (8) Exposure times before and after flare filtering. (9) Source detection sensitivity in the $ 0.5-8 $ keV energy range. (10) Major axis of the studied elliptic regions. The orientation and shape of the regions were taken from K-band measurements (http://irsa.ipac.caltech.edu/applications/2MASS/). \\

\label{tab:list1}
\end{table*}  

Although during the operation of \textit{Chandra} large number of nearby galaxies were observed, a systematic study of the unresolved X-ray emission were only performed in a few of them \citep[e.g.][]{bogdan}. Recently, \citet{boroson11} investigated a  sample of early-type galaxies, mainly focusing on their LMXB population and  ISM content. The main motivation of the present paper is to study the unresolved emission in  a large sample of early-type and late-type galaxies in a uniform manner, thereby gaining a better insight into the importance of  various unresolved X-ray emitting components. 

Our primary goal is to measure X-ray-to-K-band luminosity ratios ($L_X/L_K$) of the unresolved emission in the $0.5-2$ keV and in the $2-10$ keV energy ranges and compare their values across morphological types and stellar masses.  In particular, we aim to investigate whether $L_X/L_K$ ratios of unresolved compact sources in the old stellar population are indeed universal as was suggested earlier by \citet{revnivtsev08}. Extending this investigation to late-type galaxies we intend to study $L_X/L_K$ ratios of a stellar population that also includes young stars and YSOs. Thus, the X-ray emission from these sources and their contribution as a function of the star-formation rate (SFR) can be studied. Furthermore our sample includes  low-mass as well as massive galaxies, hence the problem of warm ISM content in galaxies will also be addressed.  

The paper is structured as follows: in Sect. 2 we introduce the sample and in Sect. 3 we discuss the methods of the data analysis. In Sect. 4 we overview various contaminating factors polluting unresolved emission. The properties of unresolved emission  in late- and in early-type galaxies are discussed in Sect. 5. The  anomalous emission in the hard band identified in four massive galaxies in Virgo cluster is discussed in Sect. 6.  We conclude in Sect. 7.

\section{Sample selection}
Since we focus on the unresolved emission, we require at least moderately deep observations allowing us to detect and remove majority of bright X-ray binaries. From the \textit{Chandra} archive we selected full-size early-type galaxies having source detection sensitivities better than  $5\times 10^{37} \ \mathrm{erg \ s^{-1}}$. However, we excluded bright cD ellipticals from our sample, such as M87 or NGC1399, furthermore we also excluded the peculiar radio galaxy, NGC5128. 

 To avoid projection effects we only selected face-on spiral galaxies with point source detection sensitivity better than $ 10^{37} \ \mathrm{erg \ s^{-1}}$ and we also included three well observed irregular galaxies. Note, that in late-type galaxies both HMXBs and LMXBs are present, hence the overall contribution of X-ray binaries is higher. Therefore a better source detection sensitivity is required to avoid substantial pollution of the unresolved emission with X-ray binaries --   for detailed discussion see Sect. \ref{sec:xraybin}. 

Our sample   consists of 26 galaxies, which includes 11 late-type and 15  early-type systems. Major properties of the selected galaxies are listed in Table \ref{tab:list1}.

\section{Data reduction}
\subsection{\textit{Chandra}}
We analyzed all available \textit{Chandra} observations for the selected galaxies, which had exposure times longer than $2$ ks. In total, this yielded $\approx4.6$ Ms data. The data  analysis was performed with standard \textsc{CIAO} software package tools  (\textsc{CIAO 4.3}; \textsc{CALDB 4.4.2}).

For each observation the flare contaminated time intervals were filtered, excluding those intervals where the count rate deviated more than $20\%$ from the mean value.  If a galaxy was observed by multiple pointings, we projected and merged each observations to the coordinate system of the observation with the longest exposure time. The combined observed and filtered exposure times for each galaxy are  given in Table \ref{tab:list1}.  To maximize the source detection sensitivity, we ran the  \textsc{ CIAO wavdetect} tool on the unfiltered combined data set in the $0.5-8$ keV band energy range. As our particular goal was to study the unresolved X-ray emission, which demands careful removal of point sources, several parameters of the \textsc{wavdetect} tool were changed from their default values. The scales on which we searched for sources were the $\sqrt 2$ series from $1.0$ to $8.0$. The value of \textsc{eenergy} was set to $0.5$, which parameter describes the percentage of PSF energy to be encircled. We also changed the values of \textsc{maxiter} to $10$ and \textsc{iterstop} to $0.00001$. The value of sigma parameter was set to 4, thereby minimizing the contribution of residual counts from points sources to the diffuse emission. 
 We thus obtained relatively large source cells, which include a large fraction of source cells. The resulting list of sources were used to mask out point sources for further analysis.

To estimate and subtract the instrumental and sky background components we used a combination of several regions away from the galaxies for all galaxies except for M31, M51, M74, M81, M83. The angular size of these galaxies is comparable or larger than the extent of the combined Chandra image, hence a direct background subtraction is not possible. Therefore we used the ACIS ``blank-sky'' files (http://cxc.harvard.edu/contrib/maxim/acisbg/) to estimate the background. As the instrumental background components of \textit{Chandra} vary with time, we renormalized the background counts using the $10-12$ keV count rates. The computed source detection senitivities are listed in Table \ref{tab:list1}, and they refer to the $0.5-8$ keV energy range assuming Galactic column density and a power law model with slope of $\Gamma=1.56$, which is typical for LMXBs \citep{irwin03}. 

To obtain X-ray luminosities of the unresolved emission in the sample galaxies we extracted spectra using the \textsc{CIAO specextract} tool. For each galaxy we used the extraction regions given in Table \ref{tab:list1}. The X-ray energy spectra were fit with a two component model, consisting of an optically-thin thermal plasma emission model (\textsc{Mekal} model in \textsc{Xspec}) with abundances fixed at Solar value \citep{anders89} and a power law model.  The column density was fixed at the Galactic value for early-type galaxies, whereas it was fixed at $N_H=10^{21} \ \rm{cm^{-2}}$ for late-type galaxies to approximately account  for the additional intrinsic absorbtion. All fits were performed in the $0.4-7$ keV energy range. Due to the relatively low number of counts in  NGC3377 and NGC3585, the temperature of the \textsc{Mekal} model was fixed at $kT=0.40$ keV.   The $0.5-2$ and $2-10$ keV band X-ray luminosities of the unresolved emission are listed in Table \ref{tab:list2}. The meaning and significance of these values are described and discussed throughout the rest of the  paper.

\subsection{Infrared data}
We used the 2MASS K-band data \citep{jarrett03} to trace the stellar light in all galaxies but M31. In case of M31 the K-band image, provided by the 2MASS archive, suffers from a background subtraction problem and is not suitable for the analysis. Instead, we rely on the $3.6 \ \mathrm{\mu m}$  Infrared Array Camera (IRAC) \citep{fazio04} data of \textit{Spitzer Space Telescope}. To estimate the background level of these images, we used nearby regions off the galaxy. To facilitate the comparison with other galaxies, we converted the Spitzer $3.6 \ \mathrm{\mu m}$ counts to 2MASS counts. The conversion factor is $C_{3.6 \ \mathrm{\mu m}}/C_{K-band}\approx10.4$, that was obtained in the central region of M31 where the role of background is negligible.

To measure star-formation rates (SFRs) in late-type galaxies we used the  $70 \ \mathrm{\mu m}$ data of Multiband Imaging Photometer (MIPS) \citep{rieke04} onboard of \textit{Spitzer Space Telescope}. The far-infrared data was converted into total infrared luminosity based on  \citet{bavouzet08}, which value was used to compute the corresponding SFR according to \citet{bell03}.

\begin{table*}
\caption{Observed X-ray luminosities and $L_X/L_K$ ratios of the sample galaxies in the $0.5-2$  and in the $2-10$ keV energy range.}
\begin{minipage}{18cm}
\renewcommand{\arraystretch}{1.3}
\centering
\begin{tabular}{c c c c c c c c c}
\hline 
Name &  kT & $L_{\rm{0.5-2keV}}$  & $L_{\rm{2-10keV}}$ & $L_{\rm{0.5-2keV},XB,sub}$ & $L_{\rm{2-10keV},XB,sub} $ & $L_{\rm{2-10keV},XB,ISM,sub} $ & $L_{\rm{0.5-2keV},XB,sub}/L_K$ & $L_{\rm{2-10keV},XB,ISM,sub}/L_K$ \\ 
     & (keV) & ($\rm{erg \ s^{-1}}$) & ($\rm{erg \ s^{-1}}$)  &   ($\rm{erg \ s^{-1}}$) &   ($\rm{erg \ s^{-1}}$) & ($\rm{erg \ s^{-1}}$)& ($\rm{erg \ s^{-1} \ L^{-1}_{K,\odot}}$)&($\rm{erg \ s^{-1} \ L^{-1}_{K,\odot}}$)\\ 
     &   (1)    &            (2)           &   (3)      &     (4)     &      (5)      & (6)        & (7)   & (8)  \\
\hline
M51        & $ 0.32 $ & $ 5.3 \times 10^{39} $ & $ 1.9 \times 10^{39} $ & $ 5.0 \times 10^{39} $ & $ 1.5 \times 10^{39} $  & $ 1.4 \times 10^{39} $  &  $ 7.9 \times 10^{28} $ & $ 2.3 \times 10^{28} $ \\   
M74        & $ 0.24 $ &  $ 4.7 \times 10^{38} $ & $ 3.5 \times 10^{38} $ & $ 4.1 \times 10^{38} $ & $ 2.3 \times 10^{38} $  & $ 2.3 \times 10^{38} $  &  $ 2.5 \times 10^{28} $ & $ 1.5 \times 10^{28} $ \\   
M81        & $ 0.32 $ &  $ 3.5 \times 10^{38} $ & $ 2.6 \times 10^{38} $ & $ 3.4 \times 10^{38} $ & $ 2.5 \times 10^{38} $  & $ 2.5 \times 10^{38} $  &  $ 7.1 \times 10^{27} $ & $ 5.2 \times 10^{27} $ \\
M83        & $ 0.34 $ &  $ 3.4 \times 10^{39} $ & $ 7.6 \times 10^{38} $ & $ 3.3 \times 10^{39} $ & $ 5.9 \times 10^{38} $  & $ 5.8 \times 10^{38} $  &  $ 8.4 \times 10^{28} $ & $ 1.5 \times 10^{28} $ \\
M94        & $ 0.47 $ &  $ 1.2 \times 10^{39} $ & $ 3.9 \times 10^{38} $ & $ 1.2 \times 10^{39} $ & $ 3.2 \times 10^{38} $  & $ 3.2 \times 10^{38} $  &  $ 3.6 \times 10^{28} $ & $ 9.6 \times 10^{27} $ \\
M95        & $ 0.24 $ &  $ 7.4 \times 10^{38} $ & $ 4.0 \times 10^{38} $ & $ 6.7 \times 10^{38} $ & $ 2.8 \times 10^{38} $  & $ 2.8 \times 10^{38} $  &  $ 2.0 \times 10^{28} $ & $ 8.5 \times 10^{27} $ \\
M101       &$ 0.22 $  &  $ 7.3 \times 10^{38} $ & $ 2.6 \times 10^{38} $ & $ 7.2 \times 10^{38} $ & $ 2.5 \times 10^{38} $  & $ 2.5 \times 10^{38} $  &  $ 2.1 \times 10^{28} $ & $ 7.4 \times 10^{27} $ \\
NGC2403    & $ 0.24 $ &  $ 7.5 \times 10^{37} $ & $ 5.1 \times 10^{37} $ & $ 7.1 \times 10^{37} $ & $ 4.4 \times 10^{37} $ & $ 4.4 \times 10^{37} $   & $ 1.4 \times 10^{28} $ & $ 8.8 \times 10^{27} $ \\
NGC3077    & $ 0.31 $ &  $ 7.8 \times 10^{37} $ & $ 6.5 \times 10^{37} $ & $ 6.2 \times 10^{37} $ & $ 4.9 \times 10^{37} $ & $ 4.9 \times 10^{37} $   & $ 2.4 \times 10^{28} $ & $ 1.9 \times 10^{28} $ \\ 
NGC4214    & $ 0.18 $ &  $ 7.5 \times 10^{37} $ & $ 1.7 \times 10^{37} $ & $ 5.1 \times 10^{37} $ & $ 1.1 \times 10^{37} $ & $ 1.1 \times 10^{37} $   & $ 7.9 \times 10^{28} $ & $ 1.7 \times 10^{28} $ \\
NGC4449    & $ 0.20 $ &  $ 5.5 \times 10^{37} $ & $ 1.6 \times 10^{38} $ & $ 5.3 \times 10^{38} $ & $ 1.4 \times 10^{38} $ & $ 1.4 \times 10^{38} $   & $ 1.6 \times 10^{29} $ & $ 4.1 \times 10^{28} $ \\
M31 bulge  & $ 0.32 $ &  $ 2.9 \times 10^{38} $ & $ 1.8 \times 10^{38} $ & $ 2.9 \times 10^{38}$  & $ 1.6 \times 10^{38}$  & $ 1.6 \times 10^{38} $   & $ 7.8 \times 10^{27}$  & $ 4.3 \times 10^{27}$ \\   
M32        & $ 0.60 $ &  $ 3.0 \times 10^{36} $ & $ 3.3 \times 10^{36} $ & $ 3.0 \times 10^{36}$  & $ 3.3 \times 10^{36}$  &  $ 3.3 \times 10^{36} $  & $ 3.5 \times 10^{27}$ & $ 3.9 \times 10^{27}$\\   
M49        & $ 0.82 $ &  $ 1.1 \times 10^{41} $ & $ 2.3 \times 10^{40} $ & $ 1.1 \times 10^{41} $ & $ 1.8 \times 10^{40} $  &  $ 1.1 \times 10^{40} $  & $ 3.6 \times 10^{29} $ & $ 3.8 \times 10^{28} $   \\
M60        & $ 0.73 $ &  $ 6.9 \times 10^{40} $ & $ 1.1 \times 10^{40} $ & $ 6.8 \times 10^{40} $ & $ 8.8 \times 10^{39} $  &  $ 5.9 \times 10^{39} $  & $ 3.1 \times 10^{29} $ & $ 2.7 \times 10^{28} $  \\
M84        & $ 0.61 $ &  $ 3.4 \times 10^{40} $ & $ 5.5 \times 10^{39} $ & $ 3.4 \times 10^{40} $ & $ 3.3 \times 10^{39} $  &  $ 2.4 \times 10^{39} $  & $ 2.1 \times 10^{29} $ & $ 1.5 \times 10^{28} $  \\
M89        & $ 0.54 $ &  $ 1.7 \times 10^{40} $ & $ 2.6 \times 10^{39} $ & $ 1.6 \times 10^{40} $ & $ 5.3 \times 10^{38} $  & $ 2.3 \times 10^{38} $  &  $ 2.3 \times 10^{29} $ & $ 3.2 \times 10^{27} $   \\
M105       &$ 0.64 $ &   $ 1.6 \times 10^{38} $ & $ 1.1 \times 10^{38} $ & $ 1.5 \times 10^{38} $ & $ 8.6 \times 10^{37} $  & $ 8.6 \times 10^{37} $  &  $ 3.7 \times 10^{27} $ & $ 2.1 \times 10^{27} $  \\
NGC1291    & $ 0.31 $ &  $ 1.3 \times 10^{39} $ & $ 4.5 \times 10^{38} $ & $ 1.3 \times 10^{39} $ & $ 2.6 \times 10^{38} $  &  $ 2.6 \times 10^{38} $  & $ 2.0 \times 10^{28} $ & $ 4.1 \times 10^{27} $  \\
NGC3377    & $ 0.4 $ &  $ 1.8 \times 10^{38} $ & $ 3.3 \times 10^{38} $ & $ 6.0 \times 10^{37} $ & $ 3.8 \times 10^{37} $  &  $ 3.8 \times 10^{37} $  & $ 3.0 \times 10^{27} $ & $ 1.9 \times 10^{27} $  \\
NGC3585    & $ 0.4 $ &  $ 3.1 \times 10^{39} $ & $ 2.4 \times 10^{39} $ & $ 2.3 \times 10^{39} $ & $ 5.4 \times 10^{38} $  & $ 5.4 \times 10^{38} $  &  $ 1.5 \times 10^{28} $ & $ 3.6 \times 10^{27} $  \\
NGC4278    & $ 0.47 $ &  $ 9.1 \times 10^{38} $ & $ 3.7 \times 10^{38} $ & $ 8.3 \times 10^{39} $ & $ 1.8 \times 10^{38} $  &  $ 1.7 \times 10^{38} $  & $ 1.5 \times 10^{28} $ & $ 3.1 \times 10^{27} $  \\ 
NGC4365    & $ 0.49 $ &  $ 3.5 \times 10^{39} $ & $ 2.0 \times 10^{39} $ & $ 2.8 \times 10^{39} $ & $ 3.4 \times 10^{38} $  &  $ 3.1 \times 10^{38} $  & $ 2.6 \times 10^{28} $ & $ 2.8 \times 10^{27} $  \\
NGC4526    & $ 0.33 $ &  $ 3.0 \times 10^{39} $ & $ 1.9 \times 10^{39} $ & $ 2.4 \times 10^{39} $ & $ 3.4 \times 10^{38} $  &  $ 3.3 \times 10^{38} $  & $ 2.5 \times 10^{28} $ & $ 3.4 \times 10^{27} $  \\
NGC4636    & $ 0.59 $ &  $ 1.2 \times 10^{41} $ & $ 6.9 \times 10^{39} $ & $ 1.2 \times 10^{41} $ & $ 5.1 \times 10^{39} $  &  $ 2.5 \times 10^{39} $  & $ 1.5 \times 10^{30} $ & $ 3.1 \times 10^{28} $  \\
NGC4697    & $ 0.33 $ &  $ 1.0 \times 10^{39} $ & $ 2.9 \times 10^{38} $ & $ 9.5 \times 10^{38} $ & $ 8.6 \times 10^{37} $  &  $ 8.3 \times 10^{37} $  & $ 1.9 \times 10^{28} $ & $ 1.6 \times 10^{27} $  \\
\hline \\
\end{tabular} 
\end{minipage}
\textit{Note.} Columns are as follows. (1) Best-fit temperature of the soft component described by \textsc{Mekal} model in \textsc{Xspec}. In case of NGC3377 and NGC3585 the temperature was fixed at the given value.  (2) and (3) Total observed unresolved X-ray luminosity in the $0.5-2$ keV and $2-10$ keV band, respectively. The  CXB contribution and the contribution of source counts falling  outside the source cells is subtracted. (4) and (5)  X-ray luminosity after the emission from the population of unresolved X-ray binaries are subtracted in the $0.5-2$ keV and $2-10$ keV band, respectively.  (6) Unresolved X-ray luminosity in the $2-10$ keV band after the emission from warm ISM is subtracted.  (7) and (8) Final, contamination subtracted, $L_X/L_K$ ratios in the  $0.5-2$ keV and $2-10$ keV band, computed from the X-ray luminosities listed in columns (4) and (6) respectively. \\
\label{tab:list2}
\end{table*}

\begin{figure*}
\hbox{
\includegraphics[width=8.5cm]{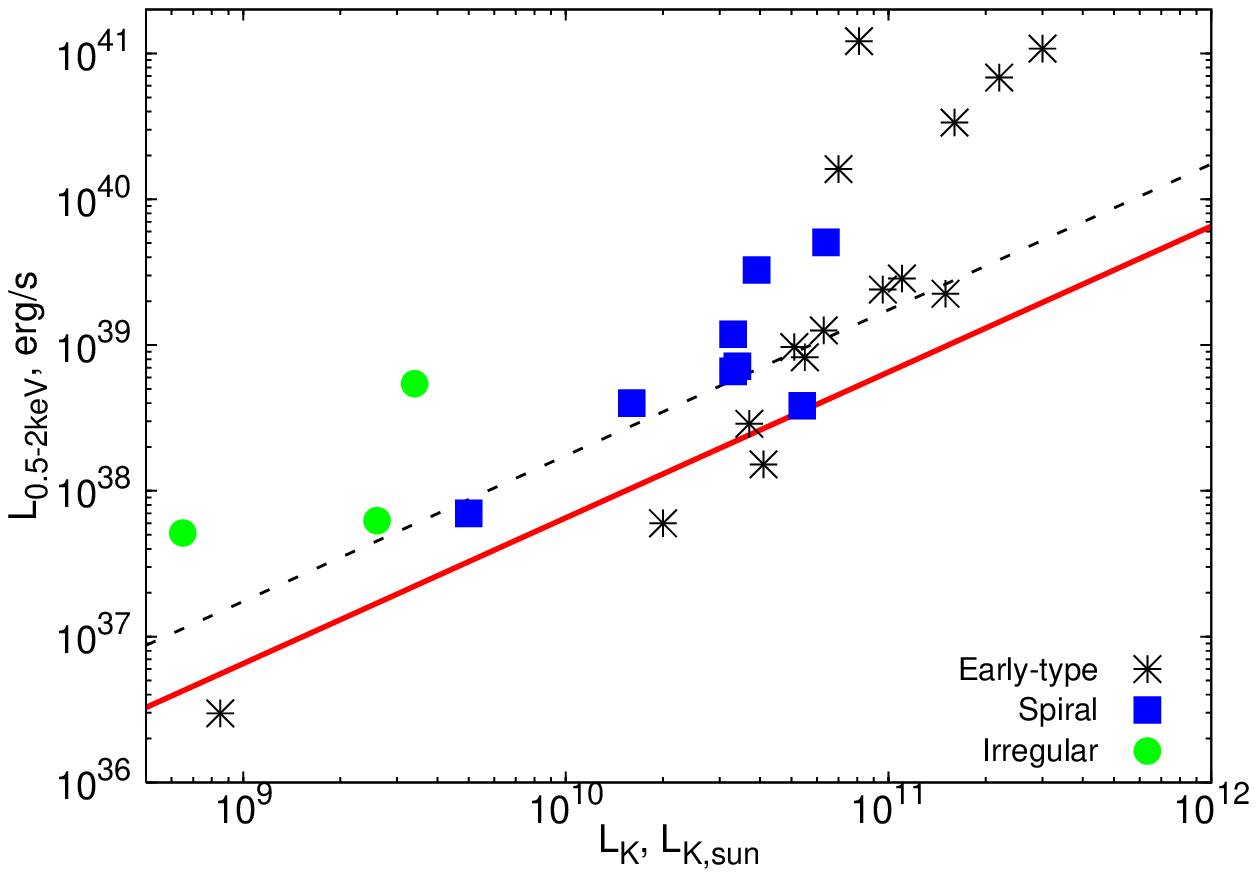}
\hspace{0.3cm}
\includegraphics[width=8.5cm]{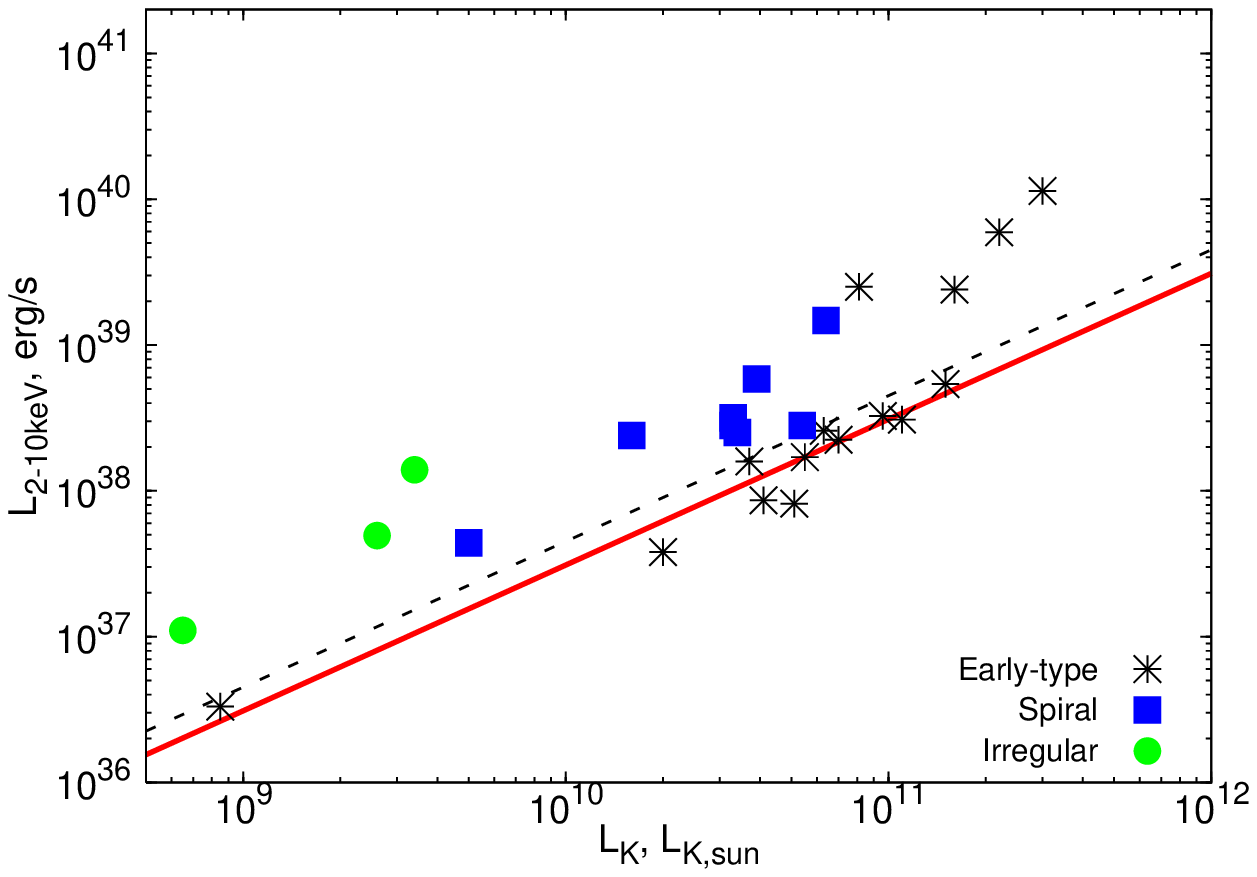}
}
\hbox{
\includegraphics[width=8.5cm]{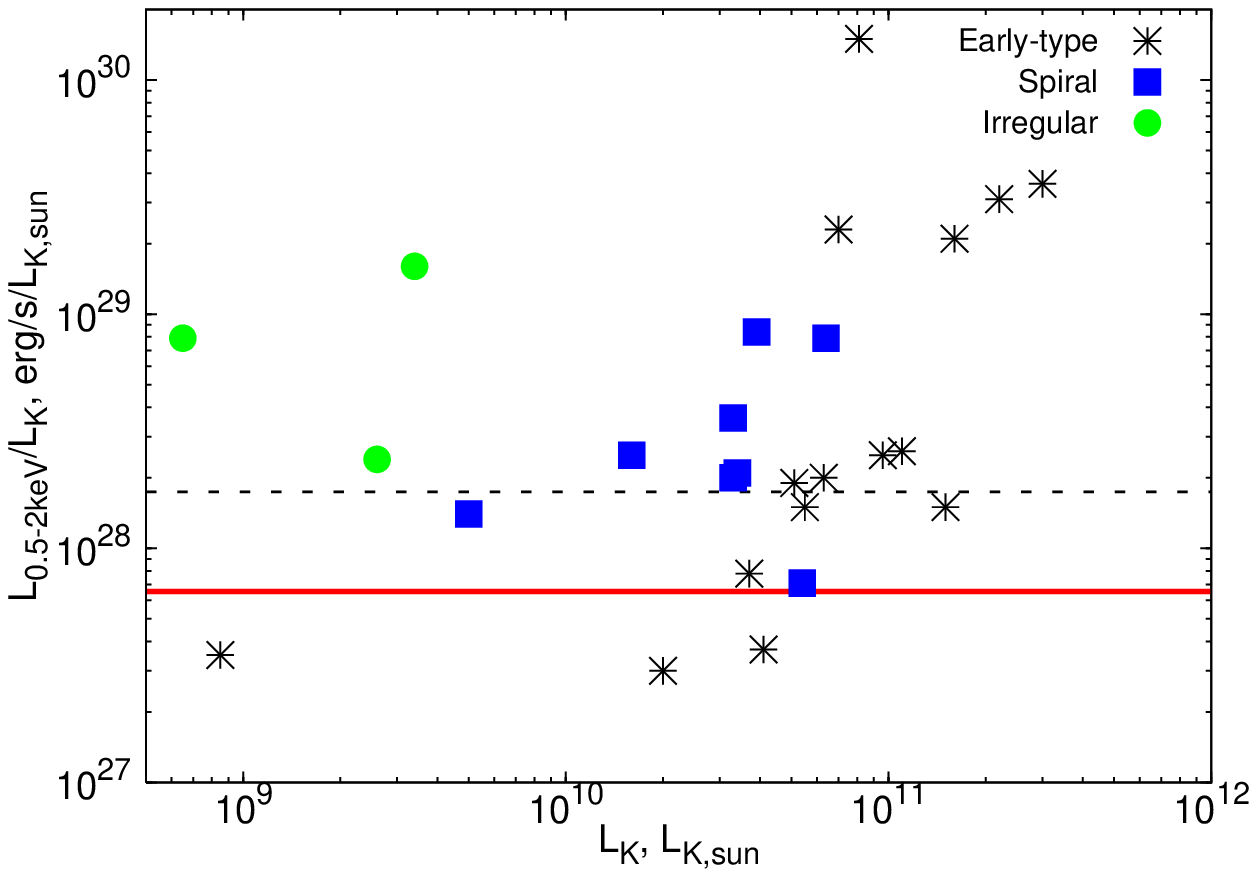}
\hspace{0.3cm}
\includegraphics[width=8.5cm]{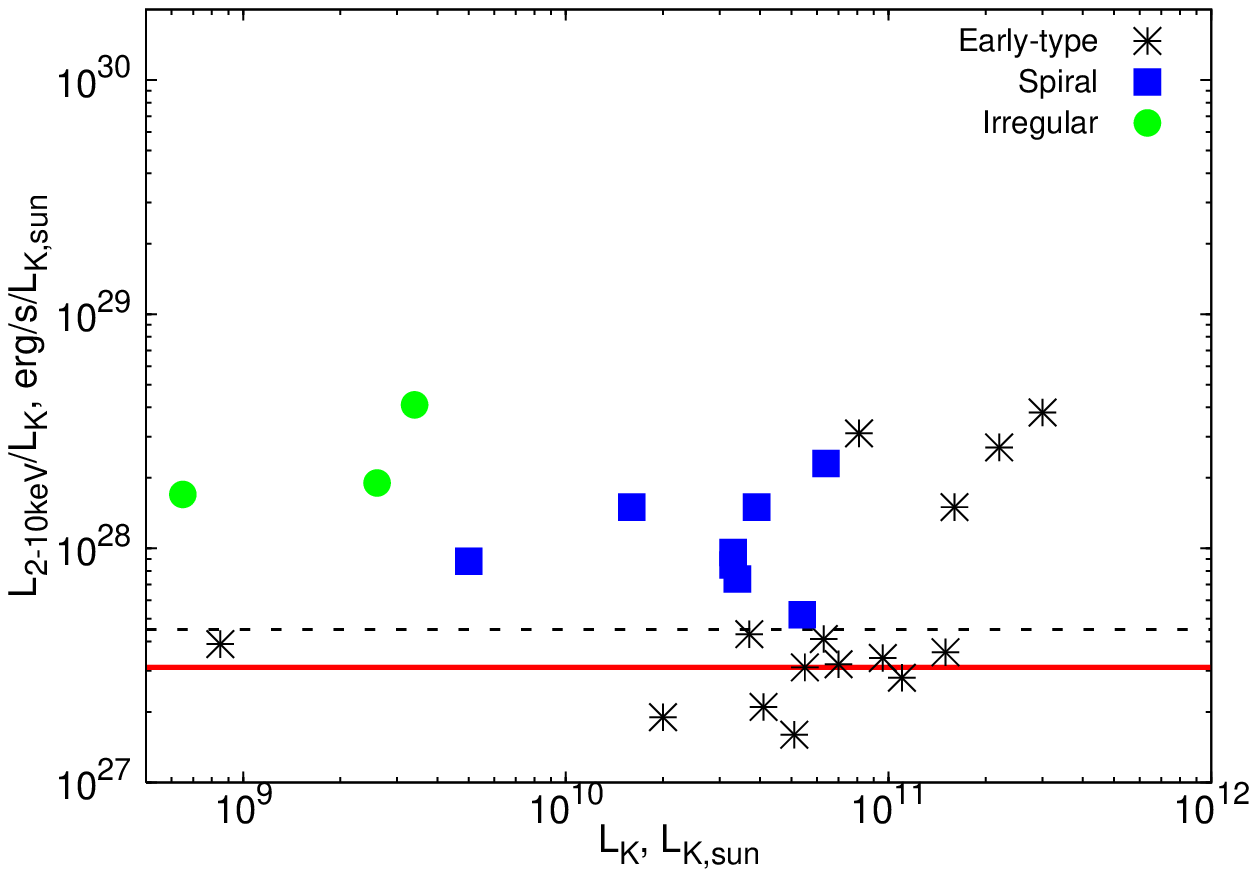}}

\caption{\textit{Top panels:} X-ray versus K-band luminosity for a set of elliptical, spiral, and irregular galaxies in the $0.5-2$ keV band (left panel) and $2-10$ keV band (right). The dashed line shows total emissivity of faint sources in the Solar neighborhood from \citet{sazonov06}. The solid line (red)  is the same but excluding contribution of  young stars. In the $0.5-2$ keV band we recalculated the \citet{sazonov06} $L_X/L_K$ ratios from the $0.1-2.4$ keV value (Sect. \ref{sec:lxlkratios}). \textit{Bottom panels:} Same data presented in the form of $L_X/L_K$ ratios. The solid and dashed lines represent the same as in the upper panels.}
\label{fig:lxlk}
\end{figure*}

\section{Sources of contamination of unresolved emission}
In the soft and hard X-ray bands a number of contaminating factors pollute the unresolved emission, whose contribution needs to be subtracted. In this  section we overview these factors, and their effect on the derived X-ray luminosities.

\subsection{Residual counts  from resolved compact sources}
\label{sec:spillover}
Although we use relatively large source cells to exclude the resolved sources, a certain fraction of source counts  falls outside these regions. Their contribution must be removed when studying unresolved X-ray emission. 

To calculate the residual emission from resolved sources we employed the following procedure. For each source we  extracted the point spread function using \textsc{CIAO mkpsf} tool and computed the fraction of counts falling outside the source cell. For most of the sources this fraction is $\sim$$ 2\%$, if it was larger, the source cell was enlarged accordingly. Few galaxies in our sample, for example NGC4278 or M81, hosts extremely bright central sources ($L_X \gtrsim 10^{40} \ \rm{erg \ s^{-1}}$), which could significantly influence the observed unresolved X-ray emission. In these cases, the source cells were enlarged to contain $>$$99\%$ of the source counts.  The residual counts from all sources were summed and subtracted from the unresolved emission. Their contribution varied between $1-25\%$.

\subsection{Unresolved X-ray binaries}
\label{sec:xraybin}

For the purpose of this study we consider all X-ray binaries, including unresolved ones,  as a source of contamination.  Resolved point sources account for the bulk of the emission from X-ray binaries and their removal is rather straightforward. They were masked out and excluded from the further analysis  as a part of the data preparation procedure. The issue of the ``spill-over'' counts is addressed in the previous section.  Unresolved X-ray binaries, on the contrary, cannot be removed on the source-by-source basis. As they make a notable contribution to unresolved emission in some galaxies, their contribution has to be removed statistically, based on the knowledge of their luminosity distributions and $L_X/L_K$ and $L_X/$SFR ratios. Therefore, we distinguish high-mass and low-mass X-ray binaries. The task is further complicated by the fact that  $L_X/L_K$ and $L_X/$SFR ratios for LMXBs and HMXBs are not exactly constant and may vary from galaxy to galaxy. 

In early-type galaxies, which host only LMXBs, we circumvent this difficulty by determining the normalization of the X-ray luminosity function (XLF) from the number of resolved X-ray binaries. In this computation we took into account that certain fraction of detected sources are not X-ray binaries but  resolved  cosmic X-ray background (CXB) sources. We computed their number based on \citet{moretti03}, and found that in all galaxies LMXBs comprise the majority,  $\approx$$ 80-97\%$, of compact sources.  We also assumed that the shape of the XLF of LMXBs is described by their average luminosity function derived in  \citet{gilfanov04}.  To transform  it to the $0.5-2$ keV and $2-10$ keV energy bands we used the average LMXB spectrum, namely  a power law with slope of $\Gamma=1.56$ \citep{irwin03}. To compute the luminosity of unresolved binaries, the luminosity functions were integrated down to $10^{35} \ \mathrm{erg \ s^{-1}}$. The thus obtained  luminosity of unresolved LMXBs  was subtracted  from the observed luminosity of each galaxy  (Table \ref{tab:list2}). 

In most galaxies the contribution of LMXBs in the $0.5-2$ keV band does not exceed $20\%$,  except for NGC3377 (67\%) and NGC3585 (26\%). In the $2-10$ keV band unresolved LMXBs play a more significant role, their contribution typically varying between $19-50\%$. But, in six galaxies with low $L_X/L_K$ ratios -- namely M89, NGC3377, NGC3585, NGC4365, NGC4526, NGC4697 -- a large fraction, $70-89\%$, of the total observed emission is due to unresolved LMXBs (Table \ref{tab:list2}). Therefore the accuracy of $L_X/L_K$ ratios of these galaxies may be somewhat compromised by residual emission  from unresolved LMXBs, because of the limited accuracy of the employed procedure. 

In late-type galaxies both LMXBs and HMXBs are present, which -- by means of X-ray observations -- cannot be distinguished.  For this reason, their X-ray luminosity functions cannot be renormalized individually.  We therefore emplyed the following procedure. We used the  luminosity functions of LMXBs \citep{gilfanov04} and HMXBs \citep{grimm03,shtykovskiy05,mineo11} with their  average normalizations to predict the numbers of low-mass and high-mass X-ray binaries above the detection threshold based on the K-band luminosity and SFR  of each galaxy, respectively. We then used the ratio of the detected number of compact sources (minus predicted CXB contribution) to the predicted number to compute the correction factor. This factor was used to correct normalizations of LMXB and HMXB XLFs  for each galaxy.  The number of CXB sources was predicted based on  source counts of \citet{moretti03}. The thus corrected XLFs were used to compute the luminosities of unresolved X-ray binaries. In converting the luminosities  of LMXBs to the soft and hard bands we used spectral parameters  specified above. For HMXBs we  assumed a power law model with slope of $\Gamma=2$, corresponding to the median value determined by \citet{swartz04}. As before, for HMXBs we used  $ N_{H} = 10^{21} \ \mathrm{cm^{-2}} $, in order to approximately account for the intrinsic absorption present in late-type galaxies.

Obviously, the method of correcting the XLF normalizations for HMXBs and LMXBs by the same factor is not  accurate. However, no better correction is possible, as  \textit{Chandra} data does not allow an unambiguous separation of the two types of X-ray binaries in external galaxies. On the other hand,   thanks to the good source detection sensitivity demanded in the selection of late-type galaxies, the contribution of unresolved binaries is rather small and this inaccuracy does not  significantly compromise the obtained $L_X/L_K$ ratios. Indeed, in the $0.5-2$ keV band the contribution of unresolved X-ray binaries does not exceed  $\lesssim15\%$. It is somewhat larger in the  $2-10$ keV energy range, but still remains below $\sim 35\%$ in all galaxies (Table \ref{tab:list2}).

\subsection{Contribution  of warm ISM to the hard band}
Although the warm ISM has a temperature in the $kT=0.2-0.8$ keV range and bulk of its emission comes below 2 keV, in some  gas rich galaxies it may contribute to the  hard band.
To be precise, this component should not be considered as a contaminating factor. However, as one of our aims is to study the $2-10$ keV emission from faint unresolved sources,  it is desired to subtract the contribution of warm ISM  from the observed hard band emission.

This calculation is rather straightforward. For the \textsc{Mekal} model with the temperature fixed at the best-fit value (Table \ref{tab:list2}) we compute the ($0.5-2$ keV)/($2-10$ keV) hardness ratio and use it to determine the luminosity of the warm ISM in the $2-10$ keV band.  Its contribution is virtually negligible for the majority of galaxies except for  M49, M60, M84, M89, and NGC4636 where it can account for up to $\sim$$10-30\%$  of the observed hard band emission (Table \ref{tab:list2}). Note, that the employed procedure slightly overestimates the contribution  of warm ISM to the hard band, because we used the total $0.5-2$ keV luminosity in this calculation.  However,  in these five galaxies the soft band emission is largely dominated  by the ISM contribution, therefore this inaccuracy does not influence the final result in any significant way.

\section{X-ray to K-band luminosity ratios}
\subsection{Observed $L_X/L_K$ ratios}
\label{sec:lxlkratios}
The final, contamination subtracted, $L_X/L_K$ ratios are listed in the last two columns of Table \ref{tab:list2} for the soft and hard bands, respectively. Additionally, the contamination subtracted  X-ray luminosities   as a function of K-band luminosities are depicted in Fig. \ref{fig:lxlk}, where the Solar neighborhood  $L_X-L_K$ relation \citep{sazonov06} is also shown. The dashed line shows  the correlation for all types of observed sources, including CVs, ABs, and young stars, whereas in drawing the solid line we excluded the contribution of young stars.  Note that  \citet{sazonov06} computed $L_X/L_K$ ratios in the  $0.1-2.4$ keV band,  hence we  converted their values into the $0.5-2$ keV energy range assuming a power law model with slope of $\Gamma=2$. 

Ideally,  gas-free early-type galaxies are expected to lie along the solid line in Fig. \ref{fig:lxlk}. Late-type systems, on the contrary, are not neccessarily expected to lie along the dashed line for two reasons. On the one hand, the \citet{sazonov06} relation was calibrated for the Solar neighborhood; on the other, the population of young stars is expected to correlate with the star-formation rate rather than the stellar mass (Sect. \ref{sec:latetype}).

Although there is an obvious correlation between the X-ray luminosity and stellar mass of the galaxy, there is large scatter in the $L_X/L_K$ ratio  in both energy bands and in galaxies of all morphological types. The observed scatter is smaller for low-mass galaxies, but more massive systems  ($L_K \gtrsim 5 \times 10^{10} \ \mathrm{L_{K,\odot}}$) tend to have $10-100$ times higher X-ray luminosities than expected based on their K-band luminosities.

\subsection{Early-type galaxies}
\label{sec:earlytype}

It is well known that  there is a broad correlation between the stellar mass of early-type  galaxies and their soft X-ray luminosity, which relation bears a large dispersion and is  non-linear, especially at the high-mass end \citep[e.g.][]{osullivan01}. Our results fit in this picture. The four least massive galaxies  with $L_K\lesssim4\times10^{10} \ \rm{L_{K,\odot}}$ have $L_X/L_K$ ratios consistent or even smaller than the Milky Way value, indicating their almost zero gas content. At higher galaxy mass the   $L_X/L_K$ increases, exceeding the Milky Way value by  factor of $ 3-250$ and reaching values characteristic of massive gas-rich ellipticals.

\begin{figure}
\hbox{
\includegraphics[width=8.5cm]{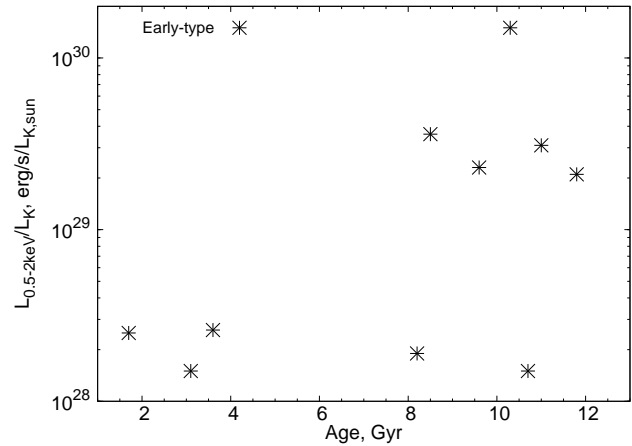}
}
\caption{Soft band $L_X/L_K$ ratios as a function of the age for gas-rich early-type galaxies  with K-band luminosity exceeding $5\times 10^{10} \ \rm{L_{\odot}}$. Note, that NGC1291 is not included since no reference for its age has been found.}
\label{fig:age}
\end{figure}

The main reason for the relatively gas-free nature of low-mass ellipticals is their shallow potential well, that permits the formation of SN Ia driven galactic-scale outflows, which are capable of removing major fraction of the ISM  \citep{david06,li07,bogdan}. Massive systems have a deeper potential well, and the energy supplied by SNe Ia is not sufficient to launch an outflow, hence the ISM will be accumulated in the galaxy. 

Calculations suggest  that an important factor influencing the gas content of a galaxy is its age \citep[e.g.][]{osullivan04},  younger galaxies having less gas. Our results appear to support this conclusion. Indeed, three rather massive galaxies at $L_K\sim10^{11}\, \rm{L_{\odot}}$, NGC3585, NGC4365, NGC4526, host fairly small amount of ISM. Their stellar ages are young: $3.1$ Gyrs \citep{terlevich02}, $3.6$ Gyrs \citep{denicolo05}, and $\sim$1.7 Gyrs \citep{gallagher08}, respectively.  On the other hand, all gas rich massive galaxies in our sample have ages of $\sim$$10$ Gyrs \citep[e.g.][]{terlevich02}. A possible existence of some correlation between $L_X/L_K$ ratio of massive galaxies with their age is further illustrated by Fig. \ref{fig:age}, where we plot all early-type galaxies with K-band luminosity exceeding $5\times 10^{10} \ \rm{L_{\odot}}$, except for NGC1291 for which no age reference has been found. The obtained correlation is not very tight, which is not surprising as a number of other parameters may play a role, such as the environment (see Sect. \ref{sec:discussion}) and the merger history of the galaxy. As another caveat, we mention that the stellar age of galaxies used for this plot were determined for central part of galaxies, typically inside  $r_e/8$ ($r_e$ is the effective radius, containing $50\%$ of the stellar light), which corresponds to  a few arcsec for the galaxies in our sample. Therefore, the listed age measurements may be inaccurate for galaxies with strong age gradients.

Three low-mass galaxies, M32, M105, NGC3377, have  $L_X/L_K$ ratios in the soft band at the $\approx (3-4)\times 10^{27} \ \rm{erg \ s^{-1} \ L_{K,\odot}^{-1}}$  level, which is by a factor of $\sim$$ 2$ lower than the Solar neighborhood value determined by \citet{sazonov06}. Two of these galaxies, M32 and NGC3377 are presumably gas-free, whereas M105 may contain a small amount of hot gas \citep{trinchieri08}. Note, that in case of NGC3377 the relatively poor source detection sensitivity may somewhat influence the obtained $L_X/L_K$ ratio (Sect. \ref{sec:xraybin}).  The observed factor of $\sim$$2$ difference may be a result of some residual contamination by young population in the Solar neighborhood value, or caused by inaccurate spectral band conversion.  In principle, it may also be a consequence of the  galaxy-to-galaxy variations, but this possibility seems less likely, in the view of the  hard band $L_X/L_K$ ratios (Fig. \ref{fig:lxlk}). 

\begin{figure}
\hbox{
\includegraphics[width=8.5cm]{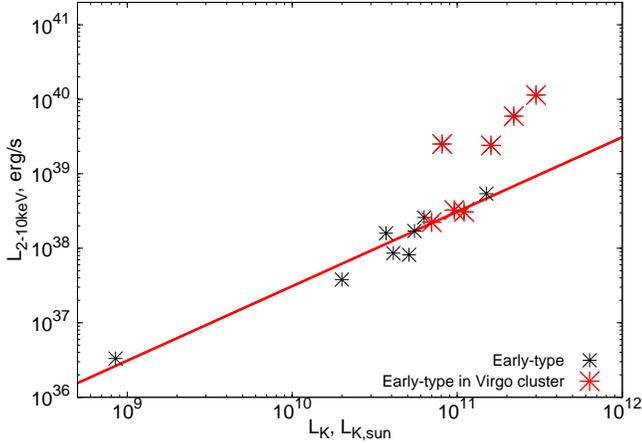}
}
\caption{X-ray luminosity in the $2-10$ keV band as a function of K-band luminosity for early-type galaxies. The contribution of warm ISM and unresolved LMXBs is subtracted. The galaxies located in Virgo cluster are marked with the large (red) symbols.}
\label{fig:early_virgo}
\end{figure}

In the $2-10$ keV energy range, the majority of early-type galaxies (11 out of 15) show fairly uniform  $L_X/L_K$ ratios,  fluctuating around  the Solar neighborhood value. The remaining four galaxies,  massive ellipticals in the Virgo galaxy cluster M49, M60, M84, and NGC4636, are significantly more luminous in the hard band. The statistical uncertainty of the $L_X/L_K$ ratio determination for these galaxies is $ (0.2-0.4) \times10^{28} \ \rm{erg \ s^{-1} \ L^{-1}_{K,\odot}} $, therefore the deviation is highly significant from the statistical point of view. Excluding these four galaxies, we calculate  for the 11 early-type galaxies  an average ratio of  $L_X/L_K=(3.1\pm0.9)\times10^{27} \ \rm{erg \ s^{-1} \ L^{-1}_{K,\odot}}$, where the cited error is the rms of the individual values. This number is in excellent agreement with the $(3.1\pm0.8)\times10^{27} \ \rm{erg \ s^{-1} \ L^{-1}_{K,\odot}}$ obtained by  \citet{sazonov06} for the Solar neighborhood. In \citet{bogdan10} we demonstrated  that the radial surface brightness profiles of gas-poor ellipticals in the soft X-ray band follow the stellar light distribution, and  that spectral charachteristics of these galaxies agree well with each other. These facts suggest that  the  bulk of the hard band X-ray emission is due to faint unresolved stellar  X-ray sources associated with old populations -- predominantly active binaries and cataclysmic variables.

Interestingly,  the four galaxies having enhanced $L_X/L_K$ ratios in the hard band, belong to Virgo cluster of galaxies (Fig. \ref{fig:early_virgo}).  The possible origin of the excess hard X-ray emission in these galaxies is discussed in Sect. \ref{sec:discussion}.

\begin{figure}
\hbox{
\includegraphics[width=8.5cm]{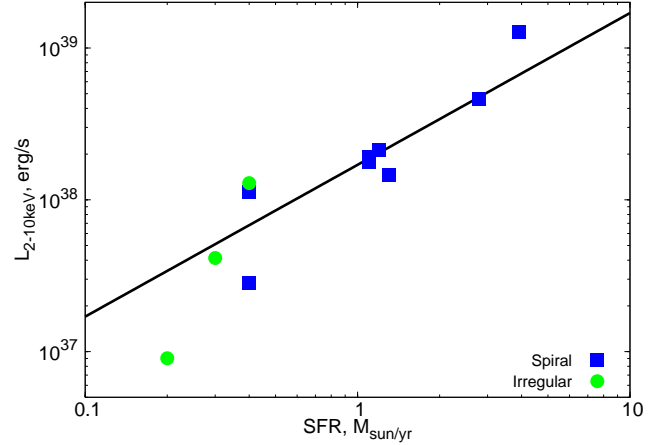}
}
\caption{Unresolved X-ray luminosity in the $2-10$ keV band as a function of the SFR for spiral and irregular galaxies. The solid line shows the relation $L_X/$SFR$=1.7\times10^{38} \ \mathrm{(erg/s)/(M_{\odot}/yr)} $ which is about $7\%$ of the total value associated with HMXBs.}
\label{fig:sfr_lxlk}
\end{figure}

\subsection{Late-type galaxies}
\label{sec:latetype}

Late-type galaxies tend to have systematically larger X-ray luminosity in both bands, than ellipticals of the same mass  (Fig. \ref{fig:lxlk}). For low-mass and medium-mass galaxies this is not too surprising, as  late-type galaxies host additional emission components -- gas and young compact sources, associated with star formation.  A similar comparison with massive ellipticals cannot be made because the mass of late-type galaxies in our sample is limited by  $L_K\sim 7\times 10^{10}~L_\odot$. The $L_X/L_K$ ratios of late-type galaxies exhibit a large scatter, not only in the soft band but also in the hard band. This can be caused by three reasons: (i) similarly to ellipticals, the gas content of late-type galaxies may vary from galaxy to galaxy; (ii) the population of young stars and YSOs must be determined by the star-formation rate rather than by the mass of the galaxy, resulting in the SFR-dependent $L_X/L_K$ ratios;  (iii)  the $L_X/L_K$ ratio of the populations of  CVs may depend on the average age of the stellar population, which may be rather  young and varying in  irregular galaxies and disks of spiral galaxies. 

The X-ray energy spectra of all late-type galaxies in our sample shows an enhanced soft component as demonstrated in \citet{bogdan11}. This component can be described by an optically-thin thermal plasma emission spectrum  with the temperature in the range of  $kT\sim 0.2-0.5$ keV (Table \ref{tab:list2}), suggesting the gaseous nature of the emission. It is plausible to conclude that all late-type galaxies contain at least moderate amount of ionized gas. Variations in the amount of hot gas  may account partly or entirely for  the observed scatter in the $L_X/L_K$ ratios in the $0.5-2$ keV energy range. The contribution of this component is negligible above $\sim$$2$ keV.

In the  $2-10$ keV band the high $L_X/L_K$ ratios are caused by the contribution of young stars and YSOs which are known to be sources of hard X-ray emission \citep{koyama96}. As the population of these sources is determined by the star-formation rate, a correlation of the hard band luminosity with the SFR is expected. In  Fig. \ref{fig:sfr_lxlk} we depict the $2-10$ keV band  luminosity of the young population as a function of the SFR. In order to remove the contribution of the X-ray sources associated with the old population,  we subtracted the luminosity corresponding to   $L_X/L_K=3.1\times10^{27} \ \rm{erg \ s^{-1} \ L^{-1}_{K,\odot}}$ from the observed values.   Fig. \ref{fig:sfr_lxlk} shows that  a correlation with SFR exists indeed, albeit with some scatter.  For our sample we calculated the average ratio of $L_X/\rm{SFR}=(1.7\pm0.9)\times10^{38} \ \mathrm{(erg/s)/(M_{\odot}/yr)}$, where the error refers to the rms of the observed values. This relation is shown in Fig. \ref{fig:sfr_lxlk} by the solid line.  We also performed a two parameter fit to the data in the form $L_X=a\times L_K+b\times \rm{SFR}$ and determined the following best-fit values for the scale factors: $a=(2.1\pm2.8)\times10^{27} \ \rm{erg \ s^{-1} \ L^{-1}_{K,\odot}}$ and $b=(2.5\pm0.6)\times10^{38} \ \mathrm{(erg/s)/(M_{\odot}/yr)}$. While the scale factor $b$ is consistent with the $L_X/$SFR value derived above, the uncertaintiy of the factor $a$  is too large to make any meaningful conclusions regarding the average $L_X/L_K$ ratio in late-type galaxies. 
Note, that the obtained $L_X/$SFR ratio for unresolved young stars and YSOs  corresponds to  $\sim$$7\%$ of the total X-ray luminosity associated with HMXBs \citep{grimm03, shtykovskiy05, mineo11}.

\begin{figure*}
\hbox{
\includegraphics[width=8.5cm]{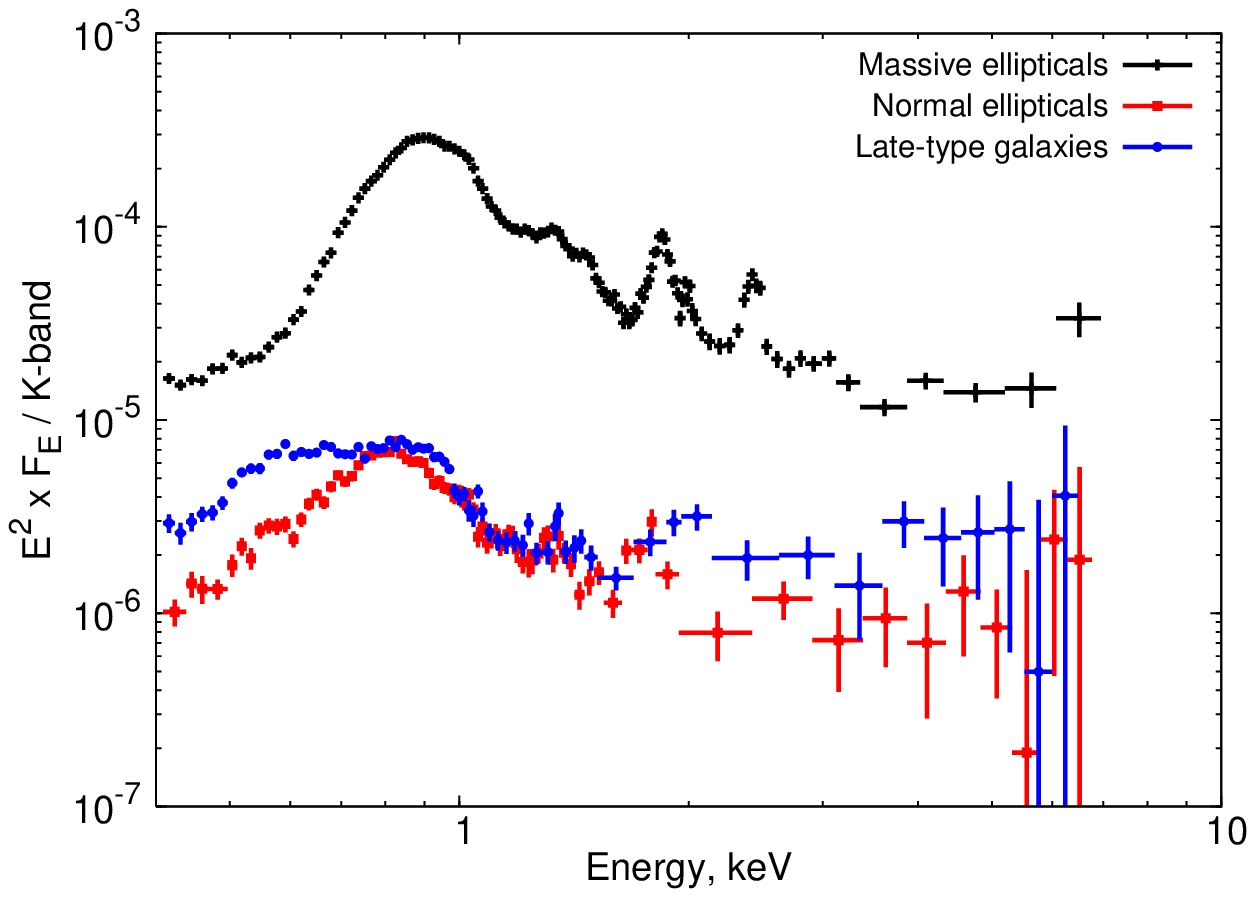}
\hspace{0.3cm}
\includegraphics[width=8.5cm]{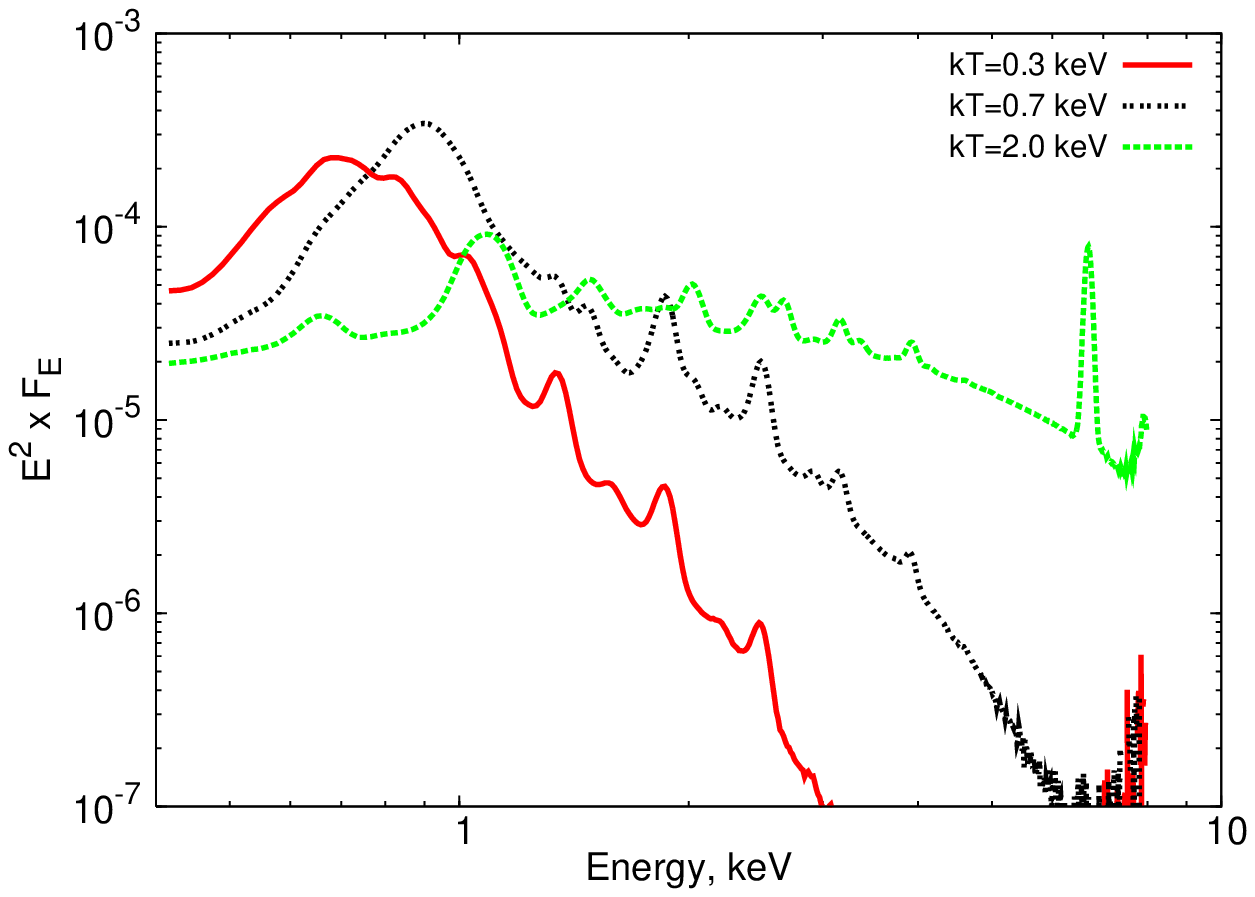}
}
\caption{\textit{Left:} Combined energy spectra of galaxies divided into three groups: 1) the galaxies with anomalous hard $L_X/L_K$ ratios -- M49, M60, M84, and NGC4636; they are  referred to as ``Massive ellipticals'' in the plot legend; 2) all remaining early-type galaxies, named ``Normal ellipticals'' and 3) all late-type galaxies. The spectra are normalized to the K-band luminosity of  $10^{11} \ \rm{L_{K,\odot}}$  and are projected to a distance of $16$ Mpc. The spectra are fluxed using the telescope efficiency for a power law spectrum with photon index of $2.09$. \textit{Right:} Optically-thin thermal plasma emission (\textsc{Mekal}) model spectra with three different temperatures: $kT=0.3$ keV (solid red line), $kT=0.7 $ keV (dotted black), and $kT=2$ keV (dashed green), solar abundance.}
\label{fig:allspec}
\end{figure*}

\section{Excess emission in the hard band in massive Virgo  ellipticals}
\label{sec:discussion}

A major unexpected result of our study is the detection of significantly enhanced $L_X/L_K$ ratios in the hard band  in four massive ellipticals (M49, M60, M84, NGC4636). 

Although the soft band $L_X/L_K$ ratio is known to vary significantly, due to varying hot gas content of elliptical galaxies, the typical gas temperatures even for the most massive galaxies in our sample do not exceed  $\approx 0.8$ keV, therefore the gas contribution to the $2-10$ keV luminosity is not dominant.  The $2-10$ keV emission from elliptical galaxies is believed to be determined by faint unresolved stellar sources, which numbers and total luminosities are expected to scale rather uniformly with the mass or K-band luminosity of the host galaxy. Indeed, the majority of ellipticals in our sample obey this expectation, following the $L_X\propto L_K$ law with a rather small scatter of  $\sim $$30\%$, in  contrast to the soft band luminosity. This behavior breaks down in the case of the four above mentioned galaxies, having approximately $5-12$ times larger $L_X/L_K$ ratios in the hard band.

\begin{table*}
\caption{Major properties of the Virgo galaxies in our sample.}
\begin{minipage}{18cm}
\renewcommand{\arraystretch}{1.3}
\centering
\begin{tabular}{c c c c c c c c}
\hline 
Name & $L_{\rm{2-10keV},XB,ISM,sub}/L_K$ & $L_{K,\rm{tot}}$ & Age & $ \sigma_c$ & $\Delta v_r$ & $d_{\rm{M87}}$ & $S_N$\\ 
     & ($\rm{erg \ s^{-1} \ L^{-1}_{K,\odot}}$) & $ (\rm{L_{K,\odot}}$) & (Gyrs) & ($\rm{km \ s^{-1}}$) &  ($\rm{km \ s^{-1}}$) & (\arcmin) & \\ 
     &   (1)    &            (2)           &   (3)      &     (4)     &      (5)    & (6)  & (7) \\
\hline
M49        & $ 3.8 \times 10^{28} $ & $ 3.8 \times 10^{11} $ & 8.5$^a$ & $293.8 \pm 2.8$  & $310 $ &  $264$ & $5.40\pm0.57^e$\\
M60        &  $ 2.7 \times 10^{28}$ & $ 2.9 \times 10^{11} $ &11.0$^a$ & $335.3 \pm 4.4$ & $190 $ & $195$ & $ 5.16\pm1.20^e$\\
M84        & $ 1.5 \times 10^{28} $  & $ 2.3 \times 10^{11} $ &11.8$^a$ &$283.3 \pm 2.8$ & $247 $ & $89$   & $ 5.20\pm1.45^e$\\
M89        &  $ 3.2 \times 10^{27} $ &  $ 9.8 \times 10^{10} $ &9.6$^a$ & $252.6 \pm 3.3$ & $967 $ & 72       & $ 2.82\pm0.57^e$\\
NGC4365& $ 2.8 \times 10^{27} $ & $ 1.9 \times 10^{11} $ &3.6$^b$ & $256.1 \pm 3.3$  & $64 $ & $319$   & $ 3.86\pm0.71^e$\\
NGC4526& $ 3.4 \times 10^{27} $ & $ 1.5 \times 10^{11} $ &1.7$^c$ & $263.7 \pm 18.9$ &$859 $ &286      & $1.09\pm0.33^e$\\
NGC4636&  $ 3.1 \times 10^{28} $ & $ 1.2 \times 10^{11} $ &10.3$^d$ &$203.1 \pm 3.5$ &$369 $ & 609     & $8.9\pm1.2^f$ \\
\hline \\
\end{tabular} 
\end{minipage}
\textit{Note.} Columns are as follows. (1) Contamination subtracted $L_X/L_K$ ratios in the $2-10$ keV energy range. (2) Total K-band luminosity from the 2MASS archive. (3) References are: $^a$ \citet{terlevich02} -- $^b$ \citet{denicolo05} -- $^c$ \citet{gallagher08} -- $^d$ \citet{sanchez06}. (4) Central velocity dispersion from HyperLeda catalog (http://leda.univ-lyon1.fr/). (5) Radial velocity relative to M87. (6) Distance from M87. (7) Globular cluster specific frequency. Refererences are: $^e$ \citet{peng08} -- $^f$ \citet{dirsch05}.
\label{tab:list3}
\end{table*}

\subsection{Common properties}

Main parameters of the Virgo galaxies are summarized in Table \ref{tab:list3}, listing their $L_X/L_K$ ratio, K-band luminosity, age, stellar velocity dispersion, radial velocity with respect to M87, offset from the center of M87, and globular cluster (GC) specific frequency. We use M87 to characterize the position and velocity  of the galaxy with respect to the intracluster medium. From the table, one may conclude that  galaxies with anomalous $L_X/L_K$ ratios  have several common properties:  (i) are members of the Virgo cluster of galaxies;  (ii) are  massive, their total  K-band luminosities are in the range of  $(1.2-3.8)\times10^{11} \ \rm{ L_{K,\odot}}$;  (iii) have relatively small velocity with respect to the intracluster gas $\Delta v_r\la 300-400$ km/s; (iv) are old, with the stellar age in the $8.5-11.8$ Gyrs range \citep{terlevich02,sanchez06}. 

Based on our, admittedly limited, sample, it appears that each of these  properties is essential. 
Indeed, the following galaxies, missing one of the above properties, have perfectly normal $L_X/L_K$ ratios: (i) a similarly massive and old Virgo galaxy M89, but having a large radial velocity with respect to M87, $\Delta v_r\approx 967$ km/s; (ii) a massive field galaxy NGC3585 with a total mass of $ 1.5\times 10^{11} \ \rm{M_\odot}$ and relatively young age; (iii) two massive, $L_K\sim 2\times10^{11} \ \rm{L_\odot}$, but apparently young Virgo galaxies NGC4365 and NGC4526 with the age of 3.6 and 1.7 Gyrs, respectively \citep{denicolo05,gallagher08}, one of them having  small ($\Delta v_r\approx 64$ km/s) and the other large ($\Delta v_r\approx 859$ km/s) velocity. Interestingly, there seems to be no dependence on the radial distance from the center of Virgo cluster, which may be explained by the somewhat irregular spatial distribution of its intracluster gas. 

As one can see from the Table \ref{tab:list3}, the four galaxies with anomalously high  $L_X/L_K$ ratios have the highest values of the globular clusters specific frequency. Therefore their LMXB populations have a larger contribution of X-ray binaries formed dynamically in globular clusters, which could, hypothetically, result in enhanced X-ray luminosity in the hard band, due to unaccounted contribution of faint sources formed in globular clusters. This is further discussed in the Section \ref{sec:faint_src}) and it is shown that GC sources cannot explain the anomalous hard emission. 

Because of the limited size of our sample, it is unclear, whether the observations presented above reflect true properties or are a result of a chance coincidence.  With this in mind, we proceed with the discussion of various mechanisms which may be responsible for the excess hard emission.

\subsection{Energy spectra}

In the left panel of Fig. \ref{fig:allspec} we show the energy spectra of all studied galaxies combined  in three groups: 1) the galaxies showing anomalous $L_X/L_K$ ratios --  M49, M60, M84, and NGC4636, 2) all remaining early-type galaxies  and 3) all late-type galaxies. To facilitate comparison, the spectra are normalized to $L_K=10^{11} \ \mathrm{L_{K,\odot}}$ and  are projected to the distance of $16$ Mpc.  Firstly, Fig. \ref{fig:allspec} shows that at every energy the spectra of the ``anomalous''  ellipticals is higher by $1-2$ orders of magnitude. Secondly, it is obvious that the enhanced hard $L_X/L_K$ ratios  are not a result of the soft component ``leaking'' into the $2-10$ keV band. Instead, it requires an additional hard component in the spectrum. Such a component, if of thermal origin, should have a temperature of about $ 2$ keV, as illustrated by the right-hand panel in Fig.\ref{fig:allspec}.

We fitted the spectrum with a model consisting of  two \textsc{Mekal} models and a power law component. The latter accounts for the population of unresolved LMXBs and has the  slope fixed at  $\Gamma=1.56$, and the normalization fixed at the value corresponding to the estimated  luminosity from unresolved LMXBs. The best-fit temperatures of the thermal components are  $kT_1=0.74\pm0.01$ keV and  $kT_2=1.75\pm0.12$ keV. The hotter component can be replaced by a power law with the photon index of $\Gamma=2.09\pm0.04$, giving similar fit quality.  In both models, the fit is statistically unacceptable, mainly because of the poor  approximation at low energies, indicating that the spectrum of the soft component is more complex.

\begin{figure*}
\hbox{
\hspace{0.75cm}
\includegraphics[width=7.75cm]{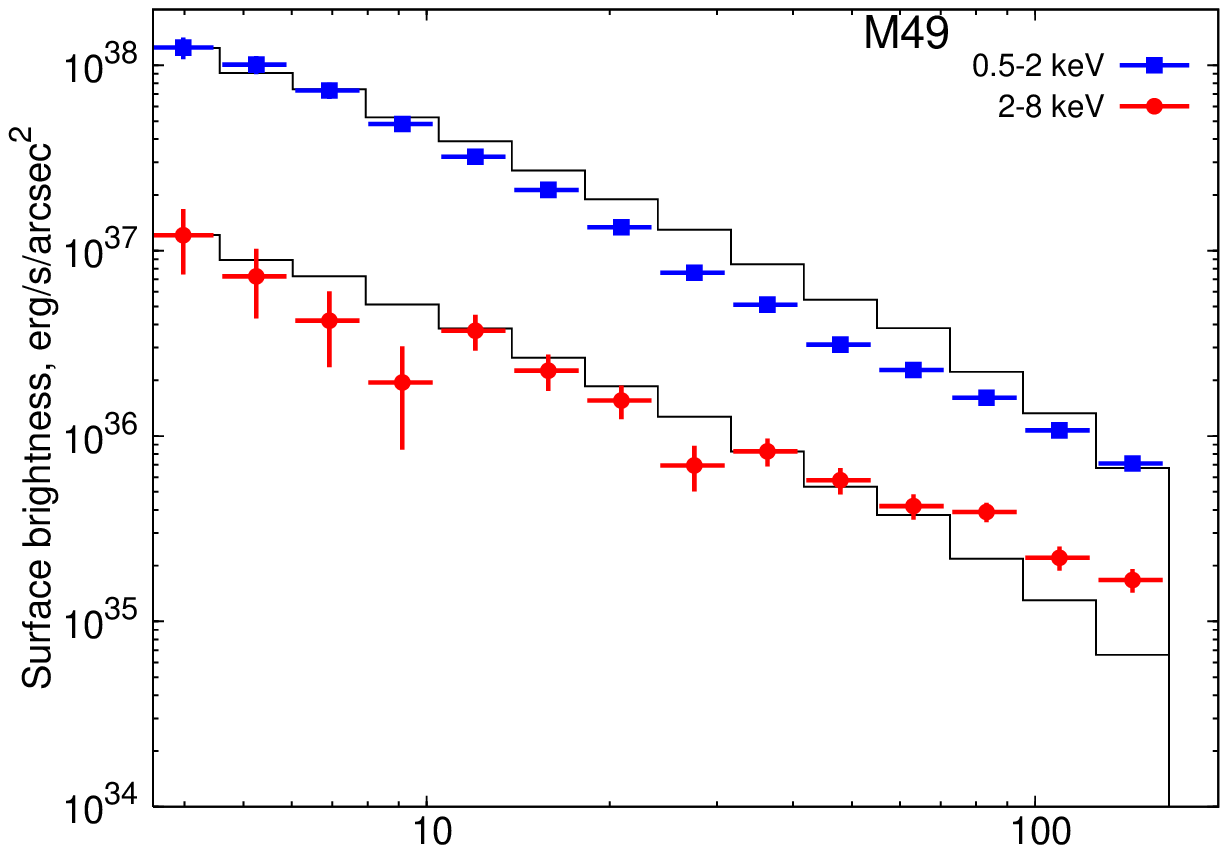}
\hspace{0.75cm}
\includegraphics[width=7.75cm]{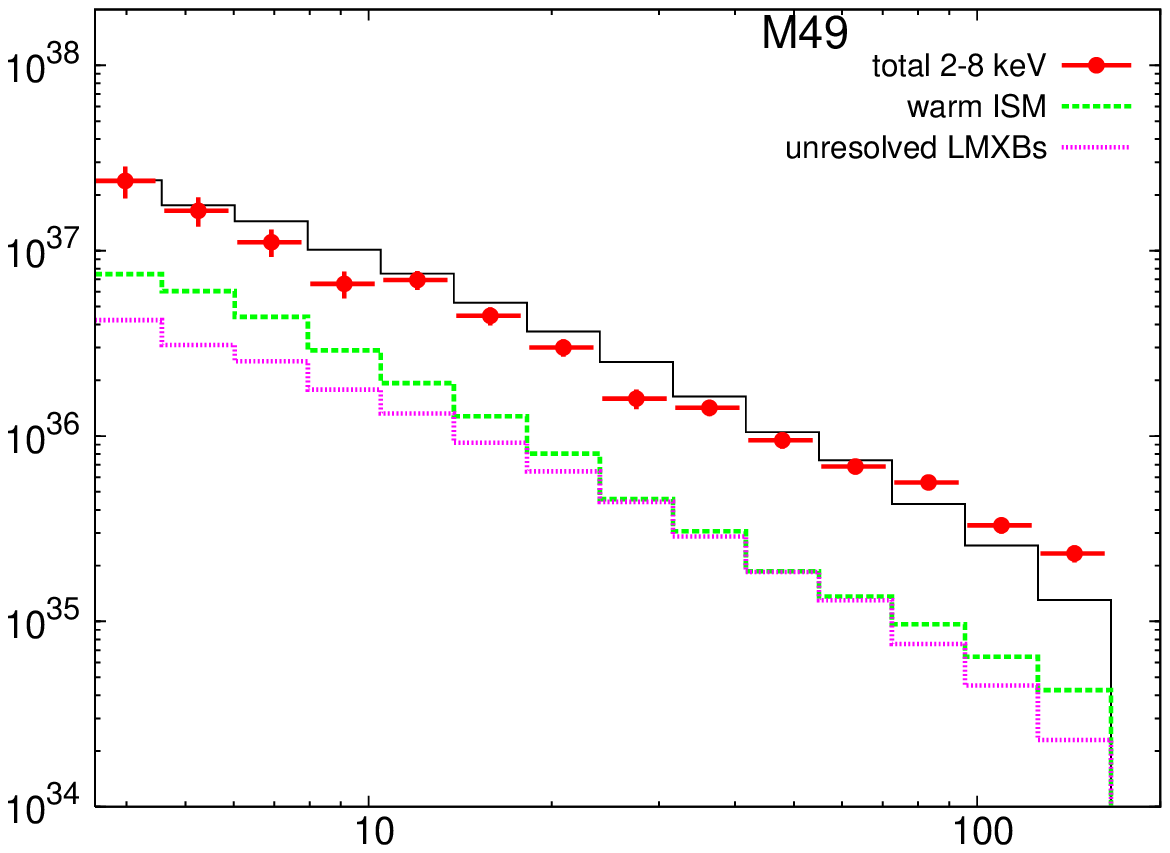}
}
\hbox{
\hspace{0.75cm}
\includegraphics[width=7.75cm]{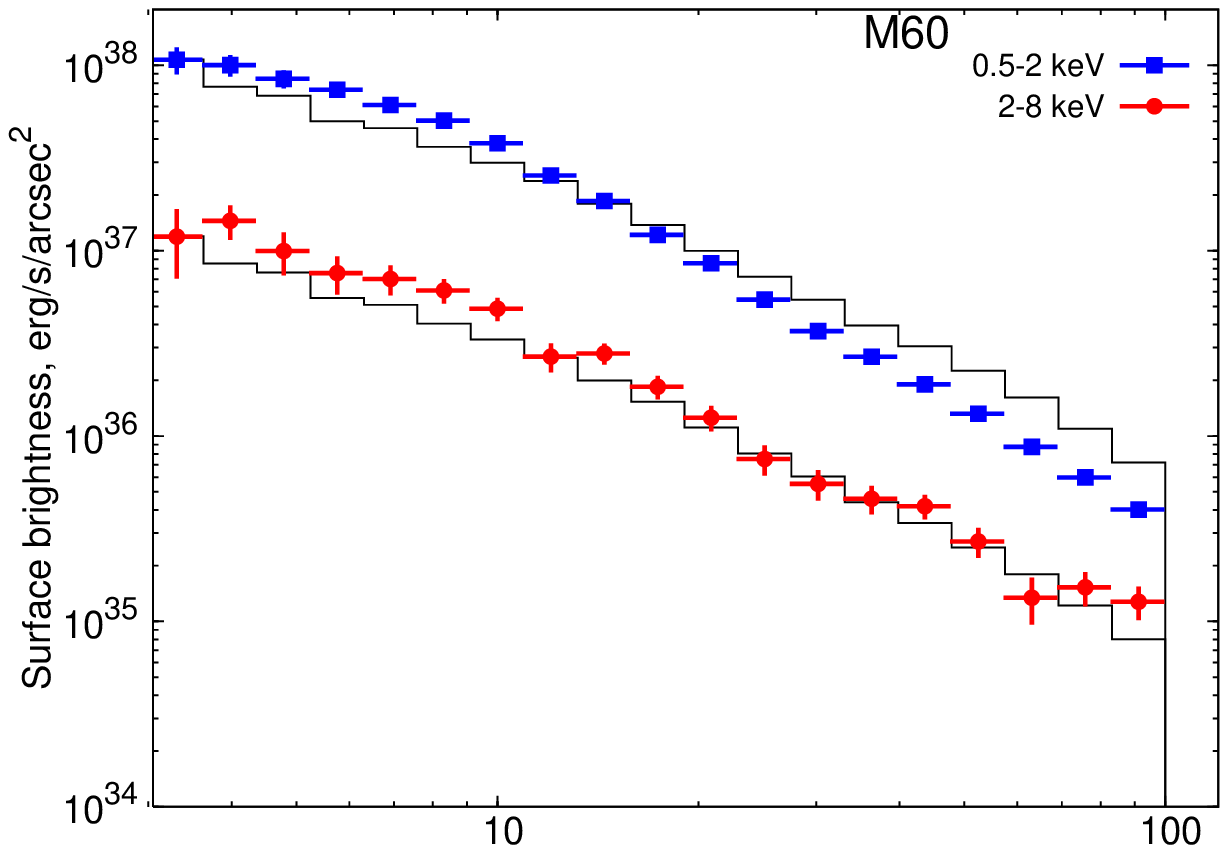}
\hspace{0.75cm}
\includegraphics[width=7.75cm]{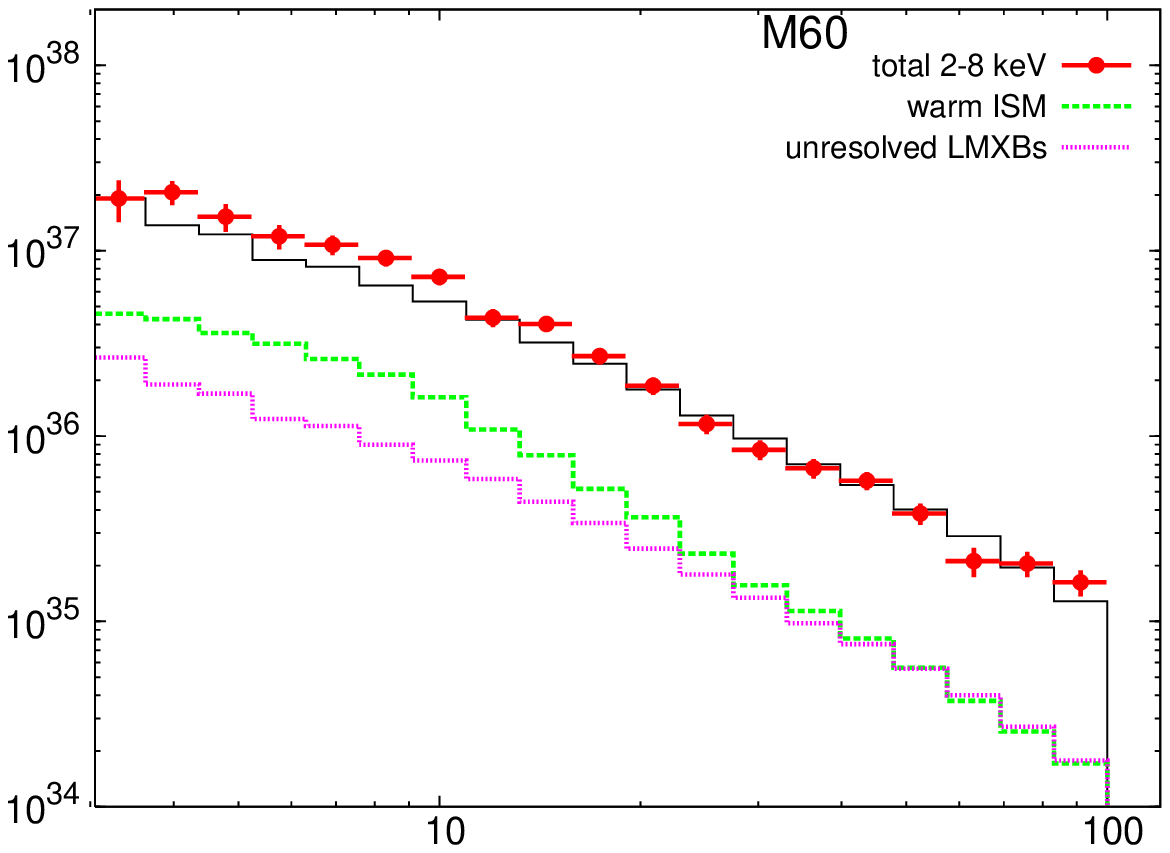}
}
\hbox{
\hspace{0.75cm}
\includegraphics[width=7.75cm]{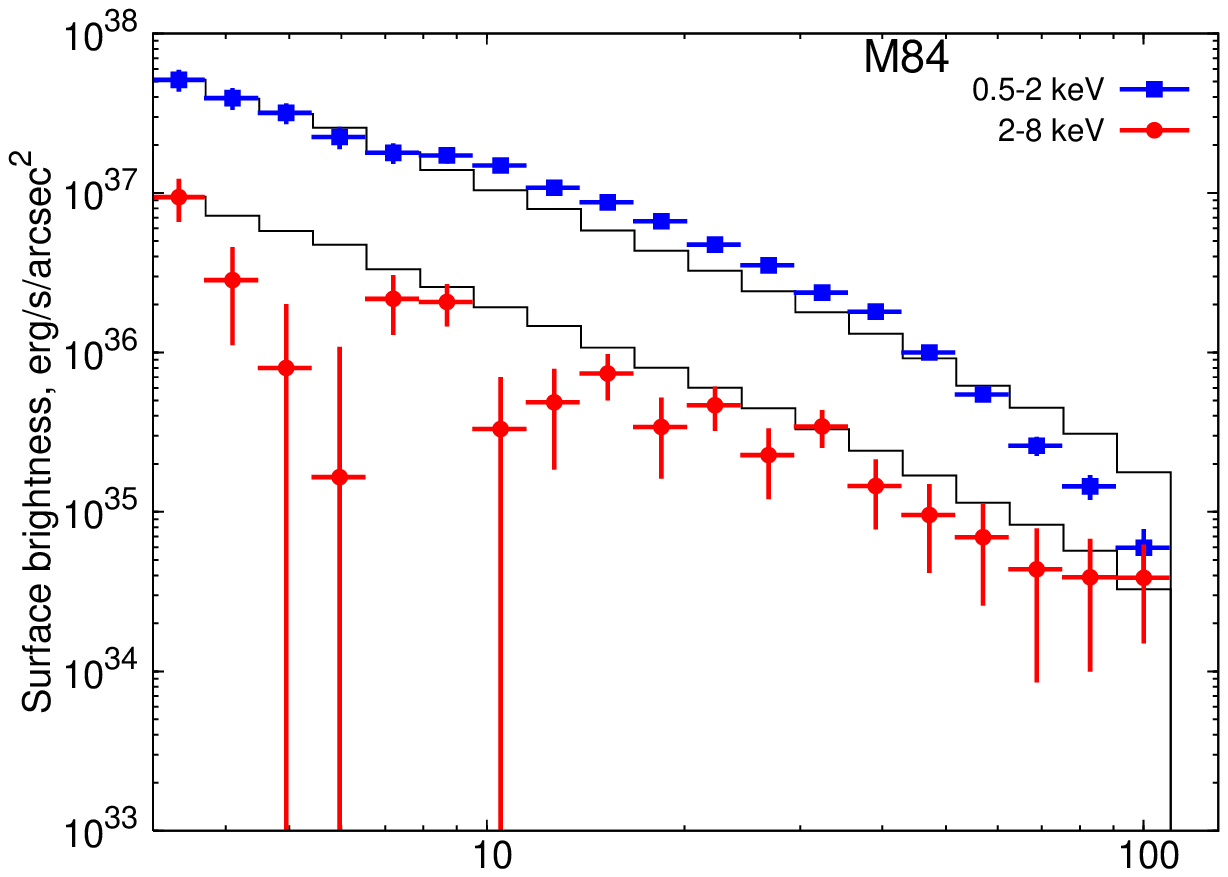}
\hspace{0.75cm}
\includegraphics[width=7.75cm]{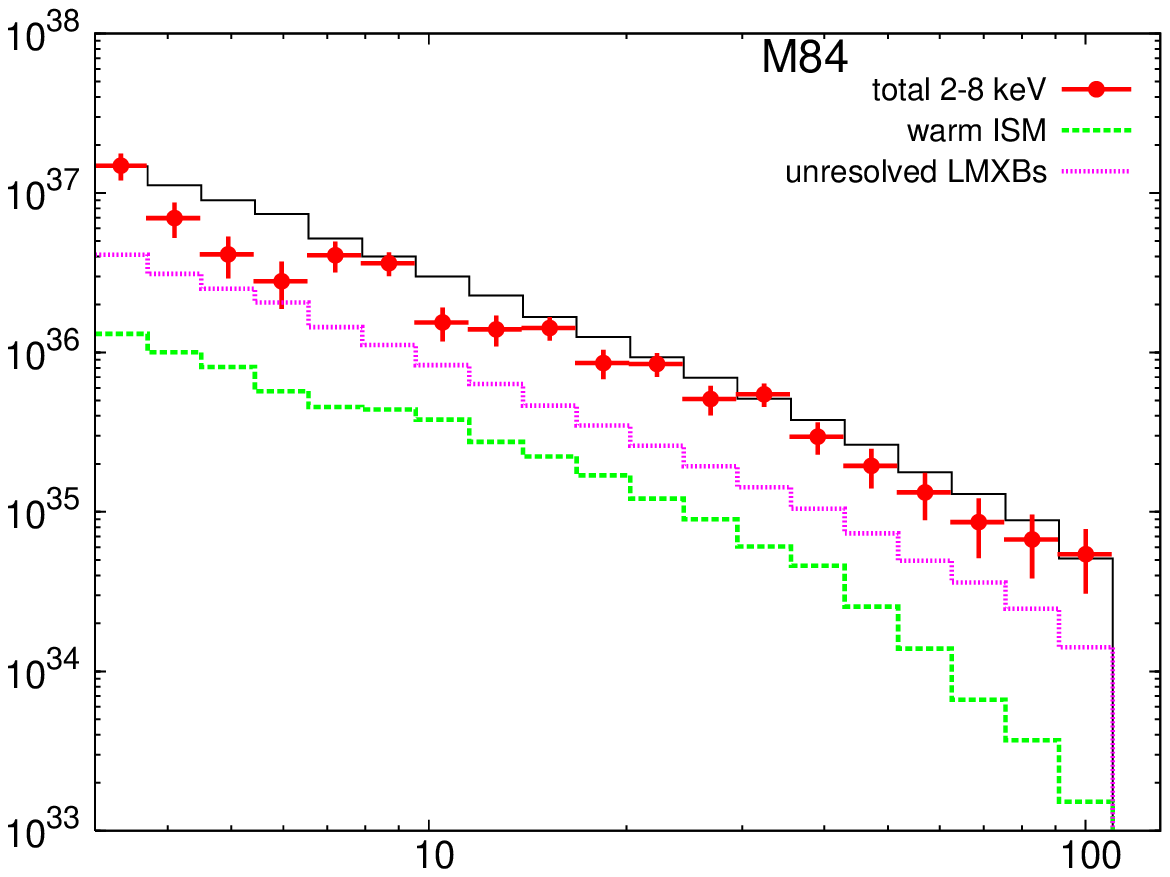}
}
\hbox{
\hspace{0.75cm}
\includegraphics[width=7.75cm]{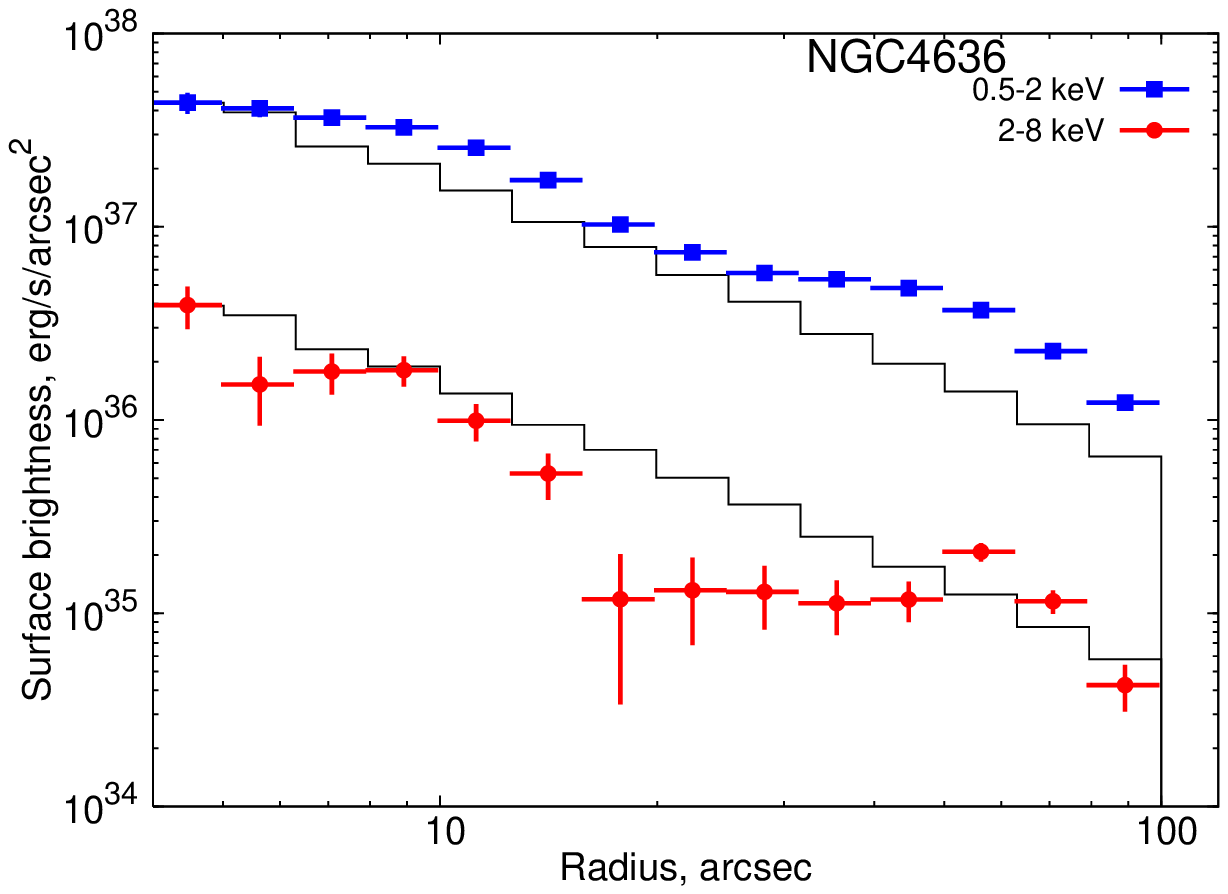}
\hspace{0.75cm}
\includegraphics[width=7.75cm]{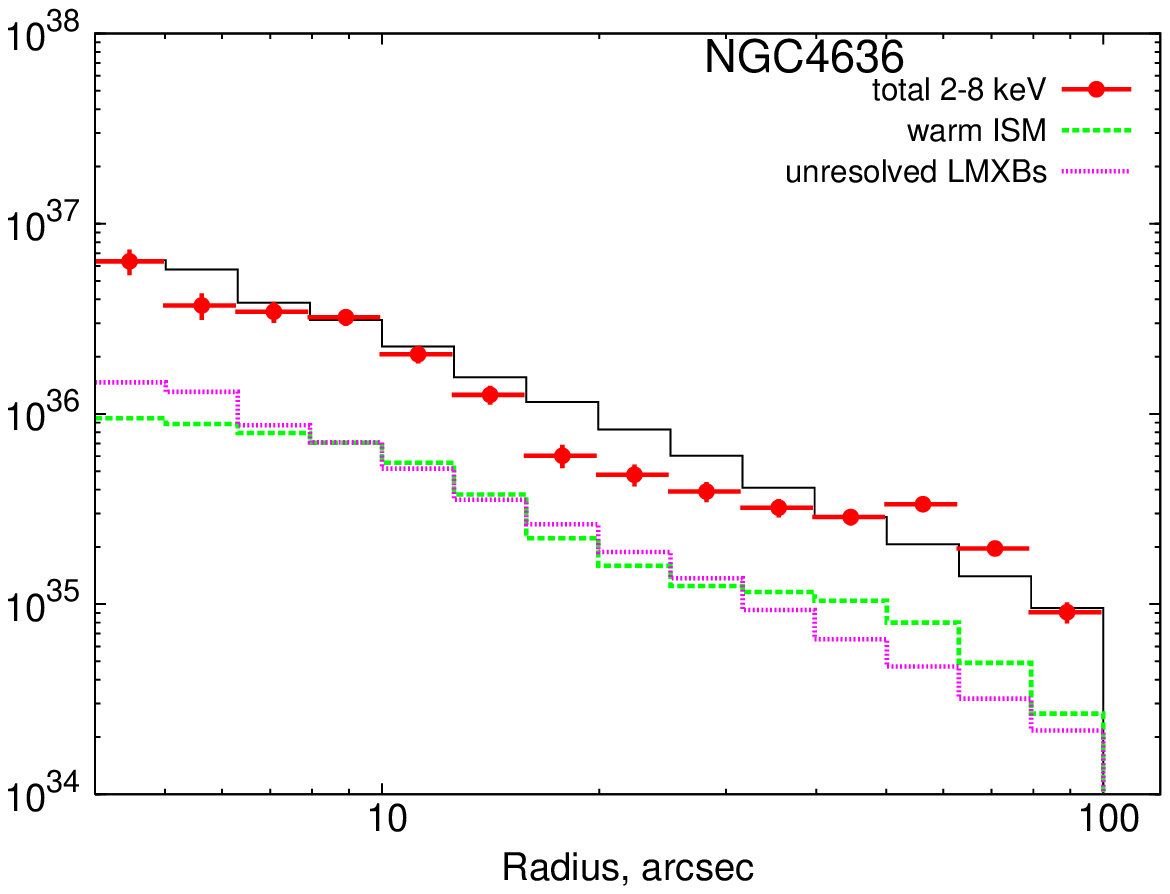}
}
\caption{\textit{Left:} Surface brightness profiles in the $0.5-2$ keV and $2-8$ keV bands in four  elliptical galaxies with anomalous hard $L_X/L_K$ ratios (data points with error bars). For each galaxy we also show the K-band light distribution (solid histogram) with arbitrary normalization.  The contaminating contribution of warm ISM and unresolved LMXBs is subtracted from the $2-8$ keV band profiles. \textit{Right:} Points with error bars show the total $2-8$ keV band X-ray surface brightness profile, before subtracting the contribution of warm ISM and unresolved LMXBs.   The radial distributions of the latter two are shown by dashed and dotted histograms.  The solid histogram shows the distribution of the K-band light.}  
\label{fig:profiles}
\end{figure*}

\subsection{Spatial distribution of hard emission}

The surface brightness profiles of unresolved emission are presented in Fig \ref{fig:profiles}. In the left four panels we show for each galaxy the soft and hard band profiles along with the K-band light distribution. The contribution of the warm ISM emission and unresolved LMXBs are subtracted from the hard band profiles. The emission from the former was computed from the soft band profiles using the best fit temperature of the soft component from Table \ref{tab:list2}. The emission from unresolved LMXBs was calculated from the K-band light distribution, the normalizations were determined as described in Sect. \ref{sec:xraybin}.  The amplitude of these two main contaminating factors is investigated  in the right hand panels, where we show the total hard band emission, before subtracting ISM and LMXB contributions  (but instrumental and cosmic background removed) and the distributions of hard band emission from ISM and unresolved LMXBs.

As one can see from the right hand panels, the hard band emission exceeds by a factor of about $5-10$ the  level of contamination (except for M84 where unresolved LMXBs contribute $\sim$$ 1/3-1/2$ of the total emission in the hard band). This is much larger than the possible uncertainty in their determination. This confirms that  the observed anomalous hard X-ray emission is not the result of these contaminating factors.   We remind that the contribution of ISM and unresolved LMXBs is subtracted from the hard band luminosities plotted in Fig. \ref{fig:lxlk} and Fig. \ref{fig:early_virgo}.

Although there are some differences between the two distributions, the soft and hard band emissions have roughly the same spatial extent, approximately similar to the spatial extent of the K-band emission.

\subsection{Possible origin of the anomalous hard emission}

\subsubsection{Accretion of hot intracluster gas}

The best fit temperature of the hard component, $kT\approx 1.7$ keV is close to the temperature of the intracluster gas in Virgo \citep[e.g.][]{bohringer94,irwin96}, suggesting that the excess emission may be due to intracluster gas accreted into the gravitational potential of a massive galaxy  \citep[e.g.][]{brighenti09}. This interpretation is qualitatively consistent with the four abovementioned properties of the galaxies with anomalous hard emission. Indeed, (i) such mechanism can work only for galaxies located in clusters; (ii) galaxies must be sufficiently massive to confine the $\sim$$ 2$ keV gas; (iii) galaxies should move sufficiently slowly with respect to the gas; (iv) some time is needed in order to accrete  sufficiently massive X-ray halo. 

One of the major facts, of such interpretation would have to explain, is the short cooling time of the gas. Given the gas density and temperature determined from the spectral fits, its cooling time inside 1 kpc does not exceed $\sim $$0.3$ Gyr in any of the four galaxies.

Another difficulty of this interpretation is  the nearly identical radial profiles in the soft and hard bands.  Indeed, in the hydrostatic equilibrium, the hotter gas component  is expected to have broader spatial distribution than the cooler one, as the scale heights for gas with $kT\approx $$1.7$ keV and $\approx$$0.7$ keV  are by factors of $\sim$$2-3$ different. This, however, is not observed -- the soft and hard band emission appear to fill the same volume.

On the other hand, the ISM may have a complex multiphase  thermal structure, with gas  of different  temperatures  present at all radii. 
Existence  of such  multiphase ISM  in massive elliptical galaxies has been proposed, for example, by  \citet{buote02} and \citet{buote03}. Based on X-ray spectroscopy of NGC1399 and NGC5044 they suggested that these galaxies host multiphase ISM. The temperature of the cooler component was found to be $kT=0.5-0.6$ keV, whereas  the temperature of the  hotter component was consistent with  the intracluster medium (ICM) temperature, i.e. similar to our results. It remains to be seen, whether accretion of hot ICM can explain anomalous hard emission from the four massive galaxies. A major advance in testing this hypothesis can be achieved by means of high resolution X-ray spectroscopy.

\subsubsection{Faint compact objects}
\label{sec:faint_src}

In principle, it is possible that the hard emission is due to an enhanced population of faint stellar sources, such as ABs and CVs. This would explain the hard spectrum of the excess emission, consistent with the power law with the photon index of $\approx $$2$ and the fact that hard band profiles approximately follow the distribution of the   K-band light. Although this assumption cannot be ruled out entirely, we stress that the numbers of resolved LMXBs in these galaxies are in very good agreement with the prediction based on the average scaling relations for LMXBs \citep{gilfanov04}. The stellar ages of the galaxies under consideration are also entirely normal for elliptical galaxies.  For these reasons, it seems implausible  that in these galaxies the ratio of $N_{\mathrm{AB,CV}}/N_{\mathrm{LMXB}}$ is $5-12$ times higher than in other early-type galaxies from our sample. 

The  high specific frequency of globular clusters (GCs) observed in M49, M60, M84, and NGC4636 may suggest, that the excess hard band emission could be due to unaccounted contribution of undetected faint sources in GCs.  These sources may be undetected, but relatively luminous LMXBs, whose luminosities fall just below the \textit{Chandra} source detection threshold, as well as much fainter sources with luminosities in the $\log(L_X)\la 31-33$ range (CVs, quiescent LMXBs, millisecond pulsars etc).  This, however, does not seem to be a viable possibility for the following reasons. The method of the XLF renormalization is based on the numbers of all observed bright LMXBs, irrespective of their association with GCs. Therefore the possibly enhanced specific frequency of LMXBs in these galaxies, due to the contribution of systems dynamically formed in GCs, is automatically taken into account by this procedure and is included in the calculation of the  luminosity of unresolved LMXBs. In fact, the shapes of the luminosity functions of field and GC LMXBs are markedly different: in the latter the fraction of faint sources ($\log(L_X)<37$) is $\sim $$4$ times smaller \citep{zhang11}.  Therefore this procedure may slightly oversubtract the contribution of undetected GC LMXBs. It is also worth to note that the specific frequency of resolved X-ray binaries in these galaxies is not much different, within a factor of $\sim$$1.5$, from other galaxies in our sample, which fact speaks against significantly enhanced population of (bright) X-ray binaries.

The possibly enhanced population of faint  sources ($\log(L_X)\la 33$) cannot be measured directly in external galaxies but their contribution can be constrained.  From \textit{Chandra} observations of GCs in the Milky Way, the
total luminosity of faint sources even in those of them with the largest encounter rates, e.g. NGC6440 and 47 Tuc \citep{pooley03} does not exceed $ {\rm few} \times 10^{33}$ erg/s. The latter number was obtained by integrating the
luminosity functions of faint sources in NGC6440 \citep{pooley02} and in  47 Tuc \citep{heinke05} and refers to the $0.5-2.5$ keV band. The numbers of globular clusters in massive ellipticals is ${\rm few} \times 10^{3}$, therefore their total luminosity in the soft band does not exceed $\sim$$ 10^{37}$ erg/s. The soft-to-hard band ratios for faint globular cluster sources are in the $\sim$$ 1-10$ range, but brighter sources tend to have softer spectra \citep{heinke05}. Thus an upper limit
on  the hard band luminosity is in the $\sim $$10^{36}-10^{37}$ erg/s range. This is about $2-3$ orders of magnitude smaller than the hard band excess luminosity detected in galaxies under consideration, $\sim$$ 10^{39}-10^{40}$ erg/s.

An even much less likely possibility is that the enhanced X-ray emission is due to population of young starts and YSO. Using the scaling relation  $L_X/$SFR$\approx1.7\times10^{38}  \ \mathrm{(erg/s)/(M_{\odot}/yr)}$ obtained in Sect. \ref{sec:latetype} we conclude that a star-formation rate of $\sim 15-60~M_\odot$/yr is required in order to  maintain $ 2-10$ keV luminosity of $L_X=(2-11)\,\times10^{39} \ \mathrm{erg \ s^{-1}}$ (Table \ref{tab:list2}). This values are appropriate for intensely star-forming galaxies and are unrealistic for ellipticals of $\approx 8.5-11.8$ Gyrs age \citep{terlevich02,sanchez06}.

\section{Conclusion}
We investigated the properties of unresolved X-ray emission  in a broad sample of nearby  early-type and late-type galaxies based on  archival \textit{Chandra} data. 
After removing the contribution of resolved and unresolved X-ray binaries we measured  $L_X/L_K$ ratios in the $0.5-2$ keV and $2-10$ keV bands and compared them with  the Solar neighborhood values. Complementing this data with the spectral and spatial  information we concluded that the unresolved X-ray emission originates from, at least, four distinct components. 
\begin{enumerate}
\item
The population of faint unresolved sources associated with old stellar population. Based on the data for elliptical galaxies we obtained a scaling relation for its $ 2-10$ keV luminosity:  
$L_X/L_K\approx 3.1 \times 10^{27} \rm{erg \ s^{-1} \ L^{-1}_{K,\odot}}$ with the rms$\approx 0.9 \times 10^{27} \rm{erg \ s^{-1} \ L^{-1}_{K,\odot}}$, which number is in excellent agreement with the value obtained by  \citet{sazonov06} for the Solar neighborhood. The $L_X/L_K$ for the soft band and for the late-type galaxies cannot be determined  unambiguously because of the contribution of  ISM and young stellar sources. For three low-mass gas poor ellipticals we obtained in the soft band $\approx $$(3-4)\times 10^{27} \ \rm{erg \ s^{-1} \ L_{K,\odot}^{-1}}$, which is by a factor of $\sim$$ 2$ smaller than the Solar neighborhood value.

\item 
In all galaxies, warm ISM with the  temperature in the  $kT\sim0.2-0.8$ keV range is present. In our sample, the gas temperatures in late-type galaxies tend to be lower than in early-types galaxies of the same mass. Similar to the results found in previous studies \cite[e.g.][]{osullivan01}, the amount and luminosity  of the gas generally scales with the stellar mass of the host galaxy,  albeit with the large scatter. The scale factor and the scatter appear to increase with the mass of the galaxy.   

\item
The population of unresolved young stars and YSOs in late-type galaxies. The X-ray emission of this component  approximately  scales  with the SFR of the host galaxy with the average  $L_X/$SFR$=(1.7\pm0.9)\times10^{38} \ \mathrm{(erg/s)/(M_{\odot}/yr)}$. 

\item
The most unexpected result of our study is the detection of anomalous emission in the $2-10$ keV band from four old and  massive Virgo ellipticals (M49, M60, M84, NGC4636). We could not offer an unambiguous explanation of this emission. A plausible interpretation may be that it is a result of accretion of the intracluster gas in the gravitational potential of the massive galaxy. In this scenario, the two component spectra on one hand and the  similarity of surface brightness profiles on the other, would point at the multiphase nature of ISM in these galaxies. High resolution X-ray spectroscopy  may shed further light on the origin of the anomalous emission.
\end{enumerate}

\bigskip
\begin{small}

\noindent
\textit{Acknowledgements.}
The authors thank Bill Forman, Christine Jones, and Ralph Kraft for critical discussions of these results and the anonymous referee for the careful reading of the manuscript and constructive comments. This research has made use of \textit{Chandra} archival data provided by the \textit{Chandra} X-ray Center. The publication makes use of software provided by the \textit{Chandra} X-ray Center (CXC) in the application package \begin{small}CIAO\end{small}. The \textit{Spitzer Space Telescope} is operated by the Jet Propulsion Laboratory, California Institute of Technology, under contract with the National Aeronautics and Space Administration. This publication makes use of data products from the Two Micron All Sky Survey, which is a joint project of the University of Massachusetts and the Infrared Processing and Analysis Center/California Institute of Technology, funded by the National Aeronautics and Space Administration and the National Science Foundation.
\end{small}

\end{document}